\documentclass{article}

 \usepackage[main, final]{neurips_2026}

\usepackage{hyperref}
\usepackage{url}

\usepackage[utf8]{inputenc} 
\usepackage[T1]{fontenc}    
\usepackage{hyperref}       
\usepackage{xurl}            
\usepackage{booktabs}       
\usepackage{amsfonts}       
\usepackage{nicefrac}       
\usepackage{microtype}      
\usepackage{xcolor}         
\usepackage{enumitem}
\usepackage{amsmath} 
\usepackage{wrapfig}
\usepackage{longtable}
\usepackage{placeins}
\usepackage{tcolorbox}
\usepackage{listings}
\DeclareUnicodeCharacter{27E8}{\ensuremath{\langle}}    
\DeclareUnicodeCharacter{27E9}{\ensuremath{\rangle}}    
\DeclareUnicodeCharacter{00B7}{\ensuremath{\cdot}}      
\DeclareUnicodeCharacter{22EF}{\ensuremath{\cdots}}     
\DeclareUnicodeCharacter{25C7}{\ensuremath{\diamond}}   
\DeclareUnicodeCharacter{21A5}{\ensuremath{\uparrow}}   
\DeclareUnicodeCharacter{207B}{\ensuremath{{}^-}}       
\DeclareUnicodeCharacter{2022}{\ensuremath{\bullet}}    
\lstdefinestyle{leanstyle}{}

\usepackage{pifont}
\newcommand{\cmark}{\textcolor{green!60!black}{\ding{51}}}

\newcommand{\bpref}[1]{\path{#1}}   

\usepackage{url}
\usepackage{graphicx}
\usepackage{booktabs}
\usepackage{multirow}
\usepackage{subcaption}
\usepackage{makecell}
\usepackage{xspace}
\usepackage{multirow}

\definecolor{zxgreen}{rgb}{0.00,0.55,0.20}
\definecolor{zxorange}{rgb}{1.00,0.45,0.00}

\newcommand{\mlf}[0]{Mathlib4\xspace}
\newcommand{\ttt}[1]{\texttt{#1}}

\newcommand{\ourmethod}[0]{\textsc{\textbf{DAPS}}\xspace}
\newcommand{\ourselector}[0]{\textsc{\ourmethod}\xspace}

\hypersetup{
    colorlinks=true,
    urlcolor=magenta,
    linkcolor={orange!60!black},     
    citecolor={cyan!60!black},    
}

\title{
\raisebox{-0.3\height}{\includegraphics[height=2.8em]{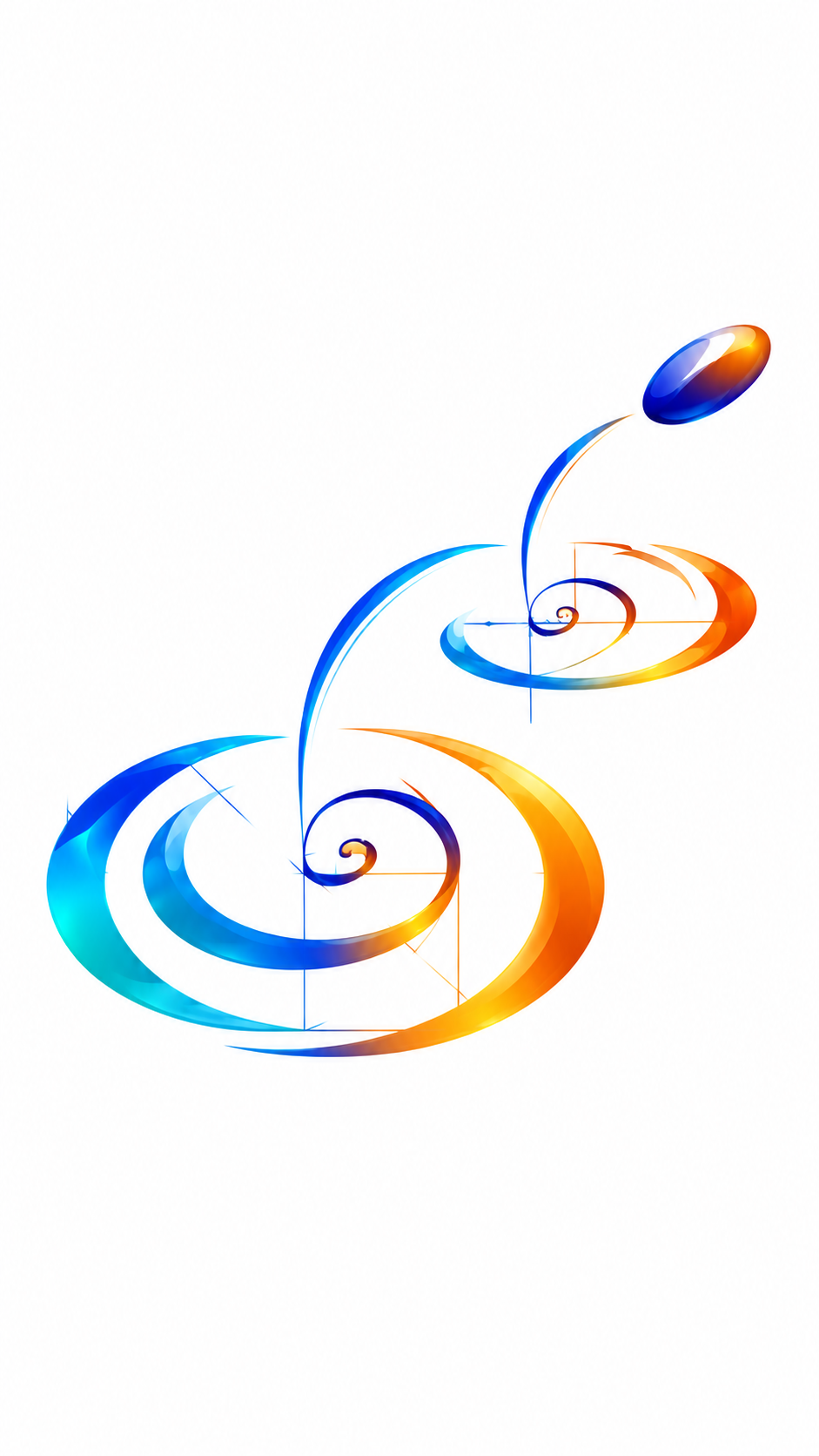}}\,DAPS: Dependency-Aware Premise Selection \\for LLM Theorem Proving
}

\author{
\textbf{Zixuan Chen*} \\
ETH Zurich \\
\texttt{zxclara@outlook.com}
\And
\textbf{Wenyuan Jiang} \\
ETH Zurich \\
\texttt{wenyjiang@student.ethz.ch}
\And
\textbf{Junling Wang} \\
ETH Zurich, ETH AI Center \\
\texttt{junling.wang@ai.ethz.ch}
\AND
\textbf{Mrinmaya Sachan} \\
ETH Zurich \\
\texttt{mrinmaya.sachan@inf.ethz.ch}
\And
\textbf{Yinya Huang*}\textsuperscript{\dag} \\
ETH Zurich, ETH AI Center \\
\texttt{yinya.huang@hotmail.com} 
}

\begin{document}

\maketitle
\begingroup
\renewcommand\thefootnote{}
\footnotetext{*Equal Contribution.}
\footnotetext{\textsuperscript{\dag}Corresponding author.}
\endgroup
\setcounter{footnote}{0}

\begin{abstract}
LLM-based theorem proving in Lean 4 has advanced rapidly, but provers are difficult to leverage newly contributed lemmas, because retrieval skews toward popular foundational lemmas, leaving the right premises buried among hundreds of thousands of candidates.
Existing neural selectors rank candidates by semantic similarity, with three intrinsic limitations: frequency-skewed retrieval, isolated pairwise ranking, and degradation on out-of-distribution queries such as competition problems and informal language. 
Given that mathematics is structured by dependencies rather than similarities, we restore this signal to both the training data and the model.
We curate two underused dependency resources.
We conduct the first systematic mining of 53 Lean Blueprint projects, a previously untapped corpus of 3{,}805 manually curated informal+formal nodes that capture rare, research-level premises absent from Mathlib4. We further pair it with the extraction of comprehensive, typed, multi-level Mathlib4 dependencies. 
We then introduce \ourmethod, a \textbf{\underline{d}}ependency-\textbf{\underline{a}}ware \textbf{\underline{p}}remise \textbf{\underline{s}}elector with a structural neighborhood encoder, a group-level contrastive objective, and a Mathlib-then-Blueprint adaptation procedure. 
\ourselector{} reaches 78.74 Recall@16 on Mathlib4-Heldout (+11.41 over the strong selector LeanHammer) and widens the margin to $7\times$ recall on competition-style queries.
Downstream, its selected premises raise Goedel-Prover-V2-8B Pass@4 on miniF2F-test from 49.90 to 54.03 and improve all four general-purpose LLMs on the informal IMO-ProofBench (up to +11.4\% relative).
%
Project page: \url{https://daps-eth.github.io/}.
\end{abstract}

\section{Introduction}
\label{sec:intro}
LLM-based Lean 4 theorem proving has progressed rapidly, with frontier provers such as DeepSeek-Prover-V2~\citep{deepseekproverv2}, Kimina-Prover~\citep{wangKiminaProverPreviewLarge2025}, and Goedel-Prover-V2~\citep{linGoedelProverV2ScalingFormal2025} reaching IMO-level performance. Yet a fundamental limitation persists: these provers operate within a self-contained paradigm, generating subgoals and reusing constants they have already memorized. Provers are difficult to leverage newly contributed lemmas in Mathlib4~\citep{mathlib2020} or theorems established by domain experts in collaborative formalization projects, because they lack a reliable way to identify the appropriate premises among the hundreds of thousands of available declarations. 
Closing this gap requires premise selection that can reliably reach the rare and recently introduced lemmas, not just the popular foundational ones the prover has already memorized.

Existing neural selectors~\citep{yangLeanDojoTheoremProving2023,zhuLeanHammerPremiseSelection2025,mikulaMagnushammerTransformerBasedApproach2024} frame premise selection as semantic retrieval, a framing with three intrinsic limitations that better encoding cannot fix.
(P1) \textbf{Frequency-skewed retrieval}: rare and recently added premises with weak text signal are displaced in top-K by popular foundational lemmas the prover already controls.
(P2) \textbf{Group-level relevance}: pairwise similarity ranks each lemma in isolation, missing whether the top-K cohere as a usable proof toolkit.
(P3) \textbf{Distribution shift}: Mathlib-trained semantic features degrade on out-of-distribution queries such as competition problems, research blueprints, and informal language.
The three share a single root cause. Mathematics is structured by dependencies rather than similarities, so what makes a lemma useful is its position in the dependency graph, for example, the classes it presupposes, the lemmas it composes with, and its distance to known premises.
Restoring this structure to both the training data and the model architecture is the unifying lever we pursue.

\begin{figure}[t]
    \centering
    \includegraphics[width=\linewidth]{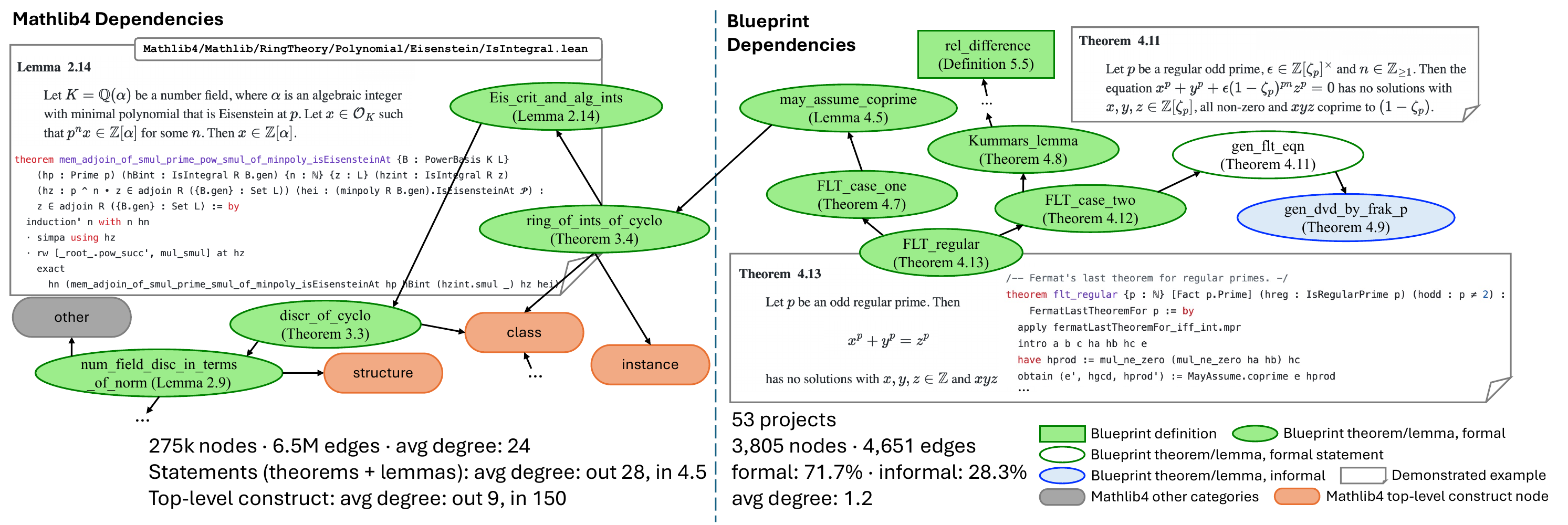}
    \caption{
        One dependency graph spanning Mathlib4 (left) and a representative Blueprint project (right).
        The two resources differ in density and composition.
        \mlf averages 24 edges per node with strong directional asymmetry across node types, while Blueprint dependencies are much sparser, averaging 1.2, but carry research-level out-of-Mathlib4 declarations and informal \LaTeX\ signals.
    }
    \label{fig:fig1}
    \vspace{-15pt}
\end{figure}

As shown in Figure~\ref{fig:fig1} with the \texttt{flt-regular}
project as a running example\footnote{\url{https://leanprover-community.github.io/flt-regular/blueprint/dep_graph_document.html}},
Lean 4 makes at least two rich dependency resources available.
\textbf{Mathlib4's compiled constant graph} (left) is a typed, heterogeneous DAG with 275k+ nodes and 6.5M edges, where each edge points from a declaration to a constant it uses in its statement or proof.
The graph is directionally asymmetric: statement nodes (e.g., \texttt{theorem}, \texttt{lemma}) have high out-degree and low in-degree, while foundational constructs (e.g., \texttt{class}, \texttt{structure}, \texttt{instance}, \texttt{inductive}) show the opposite pattern.
\textbf{Lean Blueprint projects}\footnote{\url{https://github.com/PatrickMassot/leanblueprint}} (right) are collaborative formalization repositories that contribute 3{,}805 manually curated nodes 
(71.7\% formal, 28.3\% informal) across 53 research projects. 
Many of these nodes are recent or not-yet-formalized results absent from Mathlib4, so the graph is sparse with an average degree of just 1.2.
%
\textit{Against P1's frequency-skewed retrieval}, Blueprints supply the rare, advanced premises absent from Mathlib4, while the heavy-tailed shape of Mathlib4 itself (top 10\% of nodes capture 56\% of in-edges) lets us measure the frequency skew directly.
\textit{For P2's group-level relevance}, the Mathlib graph encodes which sets of premises are jointly used per proof, providing direct supervision over premise groups rather than individual lemmas.
\textit{For P3's distribution shift}, the two resources span in-domain Mathlib (left of Figure~\ref{fig:fig1}) and out-of-distribution informal+formal queries (right), and we additionally collect out-of-distribution competition and synthetic proofs as a third source.

We accordingly derive four evaluation benchmarks of increasing distribution shift: 
\mlf-Random (1{,}000 in-distribution queries with per-declaration random holdout),
\mlf-Heldout (1{,}000 in-distribution queries with file-level isolation), 
\ourmethod-Collection (1{,}000 queries mined from 152k open-source competitions and synthetic proofs), 
and \ourmethod-Blueprint (236 informal+formal queries from advanced research projects).

We then introduce \ourmethod, a \textbf{\underline{d}}ependency-\textbf{\underline{a}}ware \textbf{\underline{p}}remise \textbf{\underline{s}}elector that addresses the three problems. 
\textit{To counter the long-tail collapse of P1},
\ourmethod has a structural neighborhood encoding that augments each candidate’s text representation with features summarizing its position in the dependency graph (i.e., the kinds of declarations it depends on, its connectivity, and its relation to the query). 
\textit{For the group-level relevance of P2},
\ourmethod has a group-level training objective that pulls the multiple gold premises of a query towards the query embedding as a single group, away from negatives, 
so the model learns to rank jointly-useful premise sets near the top, not just individual nearest neighbors.
\textit{To absorb the distribution shift of P3}, \ourmethod has a Mathlib-then-Blueprint adaptation procedure that first trains on the abundant typed Mathlib graph, then continually adapts on Blueprint dependencies.

Experiments show that \ourselector{} outperforms strong prior selectors on all four benchmarks, with the largest margins on out-of-distribution queries. 
The lead persists at small retrieval budgets on \ourmethod-Blueprint. 
Downstream, \ourselector{} premises raise Goedel-Prover-V2-8B Pass@4 on miniF2F from 49.90 to 54.03. 
They also improve four general-purpose LLMs on the informal IMO-ProofBench by up to 11.4\% over the best competing selector.


Our contributions are threefold:
\begin{itemize}[topsep=0pt, partopsep=0pt, itemsep=0pt, parsep=0pt]
\item We diagnose three intrinsic limitations of existing premise selectors: long-tail collapse (P1), group-level relevance (P2), and distribution shift (P3).
\item We bring underused Lean 4 dependency resources to bear on premise selection, specifically by systematically mining Lean Blueprint projects and comprehensive \mlf dependencies. Derived from them, we release four evaluation benchmarks of increasing distribution shift: \mlf-Random, Mathlib4-Heldout, \ourmethod-Collection, and \ourmethod-Blueprint.
\item We introduce \ourselector{}, a dependency-aware premise selector that outperforms strong prior selectors on all four benchmarks, holds its lead at small retrieval budgets, and improves downstream formal theorem proving on miniF2F and informal theorem proving on IMO-ProofBench.
\end{itemize}

\vspace{-10pt}
\section{Related Works}
\vspace{-10pt}
\textbf{Premise selection.}
Hammers~\citep{czajka2018hammerforcoq,paulsonSledgeHammerThreeYears,zhuLeanHammerPremiseSelection2025}, as a significant line of work, primarily rely on brute-force proof state suggestions, which are not well supported by LLM provers. 
Some advanced works~\citep{tworkowskiFormalPremiseSelection2022,jiangThorWieldingHammers2022a,firstBaldurWholeProofGeneration2023,mikulaMagnushammerTransformerBasedApproach2024} develop frameworks for leveraging the advantages of both hammers and LLMs, where the LLMs are trained for tool use instead of using the premises.
On the other hand, 
decomposition-based methods~\citep{jiangDraftSketchProve2023,wangLEGOPROVERNEURALTHEOREM2024,caoDSPRevivingDSP2025}
solely break down proof goals into simpler objectives or tractable subgoals, using self-generated premises.
Retrieval-augmented provers~\citep{yangLeanDojoTheoremProving2023,shenREALProverRetrievalAugmented2025} passively receive retrieved premises solely relying on semantic similarity. 
%
%
\ourmethod{} addresses this bottleneck by augmenting each candidate's text representation with its position in the Lean dependency graph and training on premise groups rather than individual lemmas, \ourmethod{} produces top-K premise sets whose collective utility is more legible to theorem proving.

\vspace{-3pt}
\textbf{Graph- and dependency-aware premise selection.}
A separate line of work treats premise selection as graph-structured retrieval over formula syntax or proof dependencies. 
Building on deep graph embeddings of formula trees~\citep{wangPremiseSelectionTheorem2017}, subsequent work strengthens the structural component via subgraph pooling~\citep{crouseImprovingGraphNeural2020},
higher-order logic encodings~\citep{paliwalGraphRepresentationsHigher2020}, 
graph contrastive pretraining~\citep{liGraphContrastivePretraining2021}, 
and graph sequence learning~\citep{holdenGraphSequenceLearning2023}.
Existing Lean dependency extractors are limited in different respects: lean-graph~\citep{leangraph} resolves precise fully-qualified names through Lean meta-programming but covers only four node types and depth-bounded subgraphs from a root, import-graph~\citep{importgraph} captures only module-level import edges and requires a successful \texttt{lake build}, and the ntp-toolkit underlying LeanHammer~\citep{zhuLeanHammerPremiseSelection2025} recovers tactic-implicit (\texttt{rw}, \texttt{simp}) references but restricts nodes to theorems and definitions.
\ourselector{} contributes a richer extraction pipeline with the broadest node and edge schema, no Lean environment or successful build required, scales to the full \mlf{} library with node-level file reconstruction, and extends to a previously untapped Blueprint corpus. It trains on this graph with a structural neighborhood encoder and group-level contrastive objective, rather than the per-formula syntax graphs and single-positive similarity ranking used by prior graph-based selectors.

\vspace{-3pt}
\textbf{LLM theorem proving.}
LLM-based theorem proving has evolved from transformer-based step
generation~\citep{gpt-f} and LLM-driven goal
decomposition~\citep{jiangDraftSketchProve2023,wangLEGOPROVERNEURALTHEOREM2024}
to large-scale RL-trained whole-proof models such as
DeepSeek-Provers~\citep{xinDeepSeekProverAdvancingTheorem2024,xinDeepSeekProverV15HarnessingProof2024,deepseekproverv2},
Kimina-Prover~\citep{wangKiminaProverPreviewLarge2025}, and
Goedel-Provers~\citep{linGoedelProverFrontierModel2025,linGoedelProverV2ScalingFormal2025},
alongside natural-language provers such as Gemini Deep
Think~\citep{luong2025advancedgeminideepthink} and
Aletheia~\citep{aletheia}. 
All operate self-contained, relying on internally generated subgoals and memorized premises, overlooking the possibility of using newly established and rare theorems from external databases.
\ourmethod{} provides the premise-side complement by surfacing rare and recently introduced lemmas from Lean libraries, enabling general-purpose LLMs to prove theorems beyond their training distribution.

\vspace{-10pt}
\section{Dependency Resources}
\label{sec:dependency-resources}
\vspace{-10pt}
We assemble two Lean 4 dependency resources, the typed dependency graph of \mlf{} (Section~\ref{sec:mathlib}) and a mined corpus of 53 open-source Lean Blueprint projects (Section~\ref{sec:blueprint-mining}), and derive four evaluation benchmarks from them (Section~\ref{sec:data}).

\vspace{-5pt}
\subsection{Mathlib Graph}
\label{sec:mathlib}
\begin{figure}[t]
  \centering
  \begin{subfigure}[t]{0.46\textwidth}
    \includegraphics[width=\linewidth]{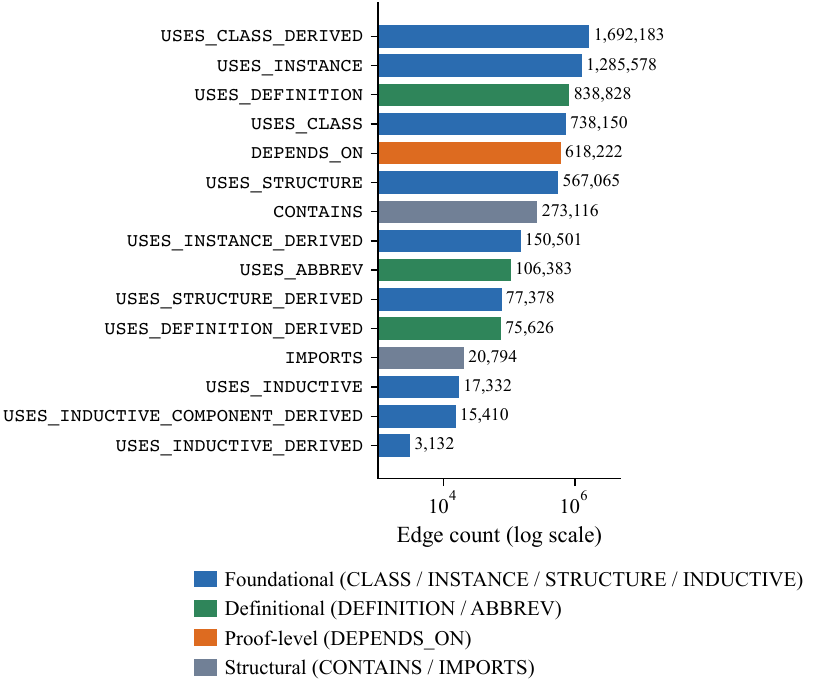}
    \caption{Typed edge spectrum.}
    \label{fig:typed_edges}
  \end{subfigure}\hfill
  \begin{subfigure}[t]{0.53\textwidth}
    \includegraphics[width=\linewidth]{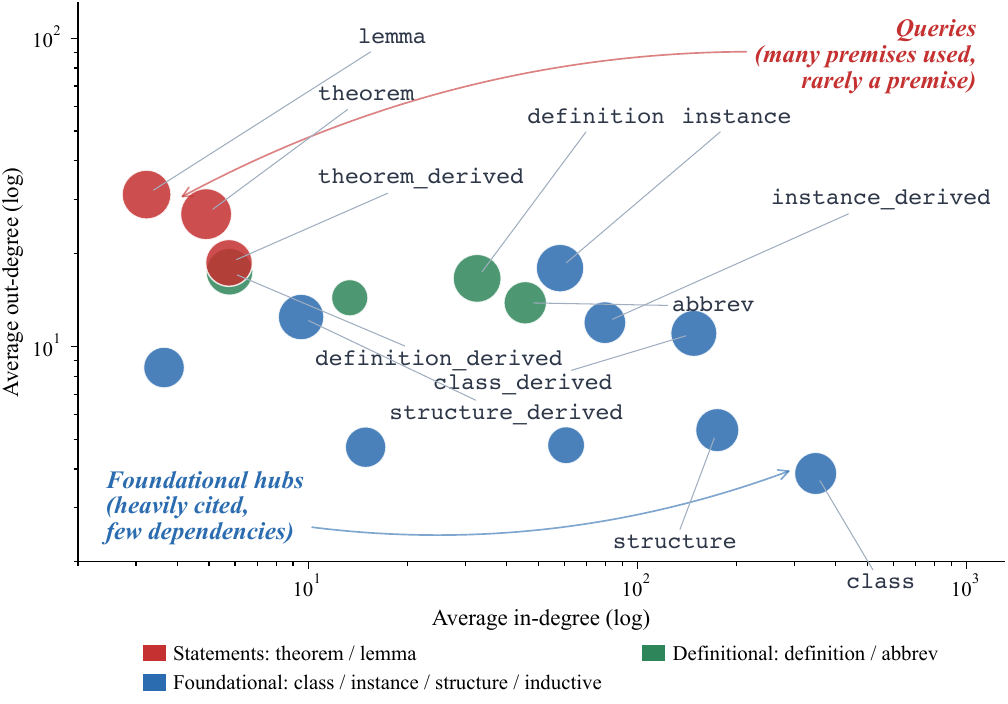}
    \caption{In-/out-degree asymmetry.}
    \label{fig:inout_asymmetry}
  \end{subfigure}\hfill
  \caption{Mathlib4's dependency graph carries a rich, asymmetric, frequency-skewed structure.
  (a) 15 typed edge categories totaling 6.5M edges.
  (b) Average in- vs.\ out-degree per node category (bubble area $\propto \log_{10}\#\text{nodes}$). Statements (red) sit at the high-out/low-in extreme, foundational type‑defining constructs (blue) at the opposite extreme. 
  }
\end{figure}

\vspace{-5pt}
We compile all Lean 4 source files in \texttt{mathlib-4.22.0}, which yields the corresponding compiled constants in the \ttt{*.olean} files for each formal declaration, including \ttt{theorems}, \ttt{lemmas}, \ttt{definitions}, \ttt{instances}, and so forth. Such compiled constants represent the dependencies.
For each formal declaration, we also record its
\texttt{type} and \texttt{docstring}.
We disable notation pretty printing and extend the constants with their fully qualified name, following LeanHammer~\citep{zhuLeanHammerPremiseSelection2025}.
Because the goal of training \ourselector is to bring together premises applicable for proving a specific theorem,
for each declaration, we further parse its statement and the whole proof from the source Lean file. We record the line number and namespace for precise positioning and reconstruction.
Additionally, we record the containment relationship between each declaration and its file, as well as the import relationship between files.
As a result, these dependencies form a multi-level directed acyclic graph, where each node is either a formal declaration or a file, 
and each edge represents a typed relationship.
Figure~\ref{fig:typed_edges} shows the edge spectrum of the constructed dependency graph. 
Appendix~\ref{appx:mathlib-graph} shows the complete node and edge categories, and further analysis.

The graph also carries strong directional structure (Figure~\ref{fig:inout_asymmetry}). Statement categories \ttt{theorem} and \ttt{lemma}, which jointly comprise 56\% of all declarations, have a low average in-degree (3--5) but a high average out-degree (27--31). These categories consume many premises but are rarely premises themselves. Foundational categories such as \ttt{class}, \ttt{structure}, and \ttt{instance} are opposite, with average in-degree up to two orders of magnitude higher and out-degree below 6. Premise selection is therefore inherently asymmetric retrieval, in which queries in the high-out, low-in end must be matched against premises in the high-in, low-out end. We encode in- and out-degree as separate features in \ourselector to capture this asymmetry.

\vspace{-5pt}
\subsection{Blueprint Mining}
\label{sec:blueprint-mining}
\vspace{-5pt}

In Lean 4, Blueprint
is a tool to manage collaborative formalization projects. 
For each project, a dependency graph is automatically generated from the human-written chapters of cross-referenced definitions, lemmas, propositions, theorems, and corollaries, which are initially written in \LaTeX\xspace and then progressively manually formalized in Lean 4. While some of these mathematical statements have been integrated into Mathlib, others have been written anew by mathematicians, and the dependencies are recorded solely in the blueprint. This is especially the case for advanced theorem formalization projects.
As a result, the blueprint dependency graphs contain rich information that is not presented in the Mathlib: 
(1) Each node has a high-quality informal statement and proof; (2) The graph contains infrequent nodes out of Mathlib. These provide high-quality and infrequent data for training \ourselector. 

We collect blueprints from open-sourced Blueprint projects. We exclude low-quality items, including those with an excessively high proportion of missing formalized statements and those with a limited number of dependencies. 
This results in 53 blueprint dependency graphs with a total of 3,805 nodes.
Table~\ref{tab:blueprint-per-project} in Appendix~\ref{appx-blueprint-list} includes detailed statistics.

\begin{table}[t]
    \centering
    \small
    \caption{Mined Lean Blueprint corpus (full per-project listing in Appendix~\ref{appx-blueprint-list}).}
    \label{tab:blueprint-summary}
    \begin{tabular}{lrrr}
    \toprule
    \# Projects                              &&& 53 \\
    \# Nodes / \# Edges                      &&& 3{,}805 / 4{,}651 \\
    Formal vs.\ informal nodes               &&& 2{,}728 (71.7\%) / 1{,}077 (28.3\%) \\
    Project size: median / mean / range      &&& 35 / 71.8 / 6--525 \\
    Fully-formalized projects                &&& 14 \\
    Informal-majority projects               &&& 14 \\
    \bottomrule
    \end{tabular}
    \vspace{-10pt}
\end{table}

\begin{table}[t]
  \centering
  \small
  \caption{
    Training and evaluation data. 
}
  \label{tab:data-splits}
  \setlength{\tabcolsep}{5pt}
  \begin{tabular}{l r r l}
    \toprule
    \textbf{Source}             & \textbf{\#Train} & \textbf{\#Val} & \textbf{Test slices (\#queries)} \\
    \midrule
    Mathlib4    & 275{,}842       & 2{,}000        & \mlf-Heldout (1{,}000), \mlf-Random (1{,}000) \\
    Blueprints   & 1{,}880 & 235           & \ourmethod-Blueprint (236) \\
    \ourmethod-Collection       & --              & --             & \ourmethod-Collection (1{,}000) \\
    \bottomrule
  \end{tabular}
  \vspace{-10pt}
\end{table}

\begin{wrapfigure}{r}{0.5\textwidth}
    \vspace{-10pt}
    \centering
    \includegraphics[width=\linewidth]{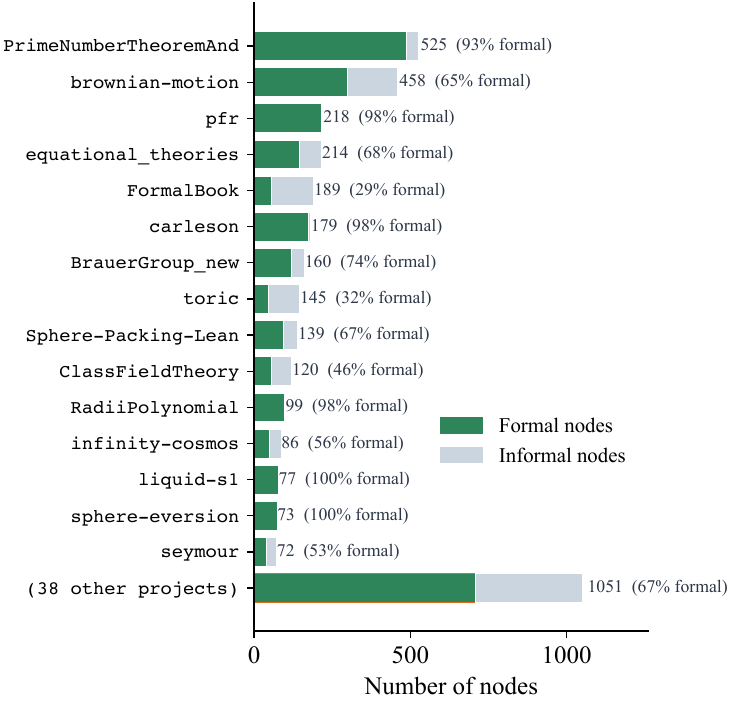}
    \caption{Composition of the mined Blueprint corpus.
  The top-15 projects are shown individually, while the remaining 38 are aggregated into the bottom bar. 
 Bars are divided into formal nodes (green; statements with Lean proofs) and informal nodes (gray; \LaTeX-only statements, sometimes with informal proofs). 
  }
  \label{fig:blueprint-composition}
  \vspace{-17pt}
\end{wrapfigure}

Table~\ref{tab:blueprint-summary} summarizes the resulting corpus, and Figure~\ref{fig:blueprint-composition} breaks it down by project. The corpus contains 3{,}805 nodes and 4{,}651 dependency edges, split 71.7\%/28.3\% between formal and informal nodes. 
Project sizes are long-tailed: the median project contains 35 nodes, the mean 72, and a handful of research-level efforts account for over half of the total, while the remaining 38 smaller projects form the low-frequency tail. 
Such research-level projects include \texttt{PrimeNumberTheoremAnd} (525 nodes), \texttt{brownian-motion} (458), \texttt{pfr} (218), \texttt{equational\_theories} (214), and the eleven FLT sub-chapters (239 combined).
Together, these projects cover number theory, analysis, algebra, topology, and combinatorics, with no single area dominating.

Crucially, the formal/informal proportion varies widely across projects. 
Per-project formal ratio ranges from 29\% (\texttt{FormalBook}) to 100\% (\texttt{sphere-eversion}, \texttt{liquid-s1}), reflecting the full life-cycle from early-stage research blueprint to completed formalization. 
14 are fully formalized, 14 remain informal-majority, and the rest sit in a mixed state. This is precisely what makes Blueprints a useful complement to Mathlib4, where every project contributes both fully-checked formal anchor nodes and informal frontier nodes that have yet to be formalized. The latter are the rare, high-value premises that no purely Mathlib-trained selector ever sees.

\vspace{-5pt}
\subsection{Data Splits and Benchmarks}
\label{sec:data}
\vspace{-5pt}
We use two training sources and one test-only out-of-distribution source (Table~\ref{tab:data-splits}).
\textbf{\mlf-Random} (1{,}000 queries) samples queries at random across all files, disjoint from the training and validation sets.
\textbf{\mlf-Heldout} (1{,}000 queries) holds out entire source files from training, a more conservative split than the theorem-level holdout used by prior work~\citep{zhuLeanHammerPremiseSelection2025}. 
\textbf{\ourmethod-Collection} (1{,}000 queries) probes query-side distribution shift. Queries are sampled from 152{,}038 \texttt{(problem, proof)} pairs aggregated from three open-source Lean~4 datasets in Olympiad and synthetic styles, with candidates still drawn from Mathlib4. 
\textbf{\ourmethod-Blueprint} (236 queries) is the test slice of an 8:1:1 split of our Blueprint corpus, stressing rare premises and informal queries. 
Full split methodology and source datasets are in Appendix~\ref{appx:data-splits}.

\section{The \ourmethod Selector}
\label{sec:method}
\vspace{-5pt}
\ourmethod is introduced to address the sketched problems Section~\ref{sec:intro} with: 
a structural neighborhood encoder for P1 (Section~\ref{sec:struct-encoding}), 
a group-level training objective for P2 (Section~\ref{sec:group-objective}), 
and a two-stage adaptation procedure for P3 (Section~\ref{sec:adaptation}). 
We first formalize the task.

\vspace{-5pt}
\subsection{Problem Formulation}
\label{sec:problem-formulation}
\vspace{-5pt}
Let $\mathcal{L} = (\mathcal{V}, \mathcal{E})$ denote a Lean library represented as a typed multi-level dependency DAG, where $\mathcal{V}$ is the set of declarations and $\mathcal{E} \subseteq \mathcal{V} \times \mathcal{V} \times \mathcal{T}$ is a typed edge set with edge categories $\mathcal{T}$ (Section~\ref{sec:mathlib}). For each query declaration $q \in \mathcal{V}$, let $\mathcal{P}^\star(q) \subseteq \mathcal{V}$ denote its \emph{positive} premises --- the declarations the verified proof of $q$ depends on. Given a candidate pool $\mathcal{C}(q)$, premise selection produces a ranking $r: \mathcal{C}(q) \to \mathbb{R}$ that places $\mathcal{P}^\star(q)$ at the top, evaluated by Recall@$K$:
\[
\mathrm{R}@K(q) \;=\; \frac{|\,\mathrm{Top}_K(r) \cap \mathcal{P}^\star(q)\,|}{|\mathcal{P}^\star(q)|}.
\]
\ourmethod instantiates $r$ by learning dependency-based embeddings $h(\cdot)$ of queries and candidates and ranking by cosine similarity $\mathrm{sim}(h(q), h(p))$.

\subsection{Structural Neighborhood Encoding}
\label{sec:struct-encoding}
\vspace{-5pt}
For each candidate $p \in \mathcal{C}(q)$, we encode both $p$ itself and a sampled neighborhood from its 2-hop dependency subgraph in $\mathcal{L}$. The neighborhood signal is what allows \ourmethod to rank rare premises whose own declaration text is short or unfamiliar (P1).

\textbf{Semantic encoding.}
A small text encoder maps the concatenation of \texttt{name}, \texttt{category}, \texttt{signature}, \texttt{docstring}, \texttt{statement}, and \texttt{proof} of any declaration $p$ to a base feature
\(h_{\mathrm{feat}}(p) = \mathrm{Encoder}(p)\).
The encoder is fine-tuned with the masked contrastive loss of \citep{zhuLeanHammerPremiseSelection2025}, with queries and premises encoded independently and their embeddings contrasted on $(q, p_{\mathrm{pos}})$ and $(q, p_{\mathrm{neg}})$ pairs, and then frozen for the rest of \ourmethod training (Appendix~\ref{appx:hyperparams}).

\textbf{Dynamic subgraph sampling.}
For each candidate $p$, we collect the 1- and 2-hop neighbors of $p$ in $\mathcal{L}$, randomly sample $m$ of them, score the $m$ candidates with a learnable two-layer FFN over $h_{\mathrm{feat}}(\cdot)$, and retain the top-$n$ scored neighbors as $\mathcal{N}(p)$. This sample-then-score selection bounds compute and memory while preferentially retaining structurally informative neighbors.
It also matches the deployment-time constraint that downstream LLM provers consume only a bounded premise list.

\textbf{Per-neighbor structural features.}
We adapt three token features from the relational graph transformer (RelGT)~\citep{dwivediRelationalGraphTransformer2025} and add one feature specific to typed dependency graphs. For each $p_i \in \{p\} \cup \mathcal{N}(p)$ we compute:
\begin{enumerate}[leftmargin=1.4em,itemsep=0.1em]
  \item \emph{Type} embedding $h_{\mathrm{type}}(p_i)$ via a one-hot over the \texttt{subcategory} field~\citep{dwivediRelationalGraphTransformer2025};
  \item \emph{Hop} embedding $h_{\mathrm{hop}}(p, p_i) \in \{0,1,2\}$ giving the dependency hop distance from $p$ to $p_i$~\citep{dwivediRelationalGraphTransformer2025};
  \item \emph{Positional} encoding $h_{\mathrm{pe}}(p_i)$ over the sampled local subgraph~\citep{dwivediRelationalGraphTransformer2025};
  \item \emph{Degree} embedding $h_{\mathrm{deg}}(p_i) = W_{\mathrm{deg}}\,[\deg^-(p_i);\deg^+(p_i)]$ of in- and out-degree, which is motivated by the directional asymmetry of the dependency graph (Section~\ref{sec:mathlib}, Figure~\ref{fig:inout_asymmetry}), in which different declaration categories occupy distinct in- vs.\ out-degree regions.
\end{enumerate}

\textbf{Fusion.}
Per-neighbor representations are merged by a two-layer MLP $O$, with the candidate's own type re-injected through a residual path so it cannot be lost in subgraph aggregation:
\[
h(p_i) \;=\; O\,\bigl[h_{\mathrm{feat}}(p_i);\, h_{\mathrm{type}}(p_i);\, h_{\mathrm{hop}}(p, p_i);\, h_{\mathrm{pe}}(p_i)\bigr] \;+\; h_{\mathrm{deg}}(p_i) \;+\; h_{\mathrm{res}}(p),
\]
where $h_{\mathrm{res}}(p) = W_{\mathrm{type}}\,\mathrm{onehot}(\phi(p))$.

\subsection{Group-Level Contrastive Objective}
\label{sec:group-objective}
\vspace{-5pt}
A useful proof typically requires several premises working together.
Therefore, unlike the typical contrastive setup that a query is pulled toward a single positive sample, \ourmethod is designed to rank a large set of candidate premises, in which multiple positives are pulled toward the query \emph{as a group} relative to the negatives (P2).

For each query $q$ in a training batch, we sample a positive set $\mathcal{P}^+_q \subseteq \mathcal{P}^\star(q)$ (the top-$\ell$ positives under the category priority of Appendix~\ref{appx:hyperparams}) and a negative set $\mathcal{P}^-_q$ via three complementary strategies, drawing on standard hard-negative mining techniques: 
(i) \emph{temporal} sampling of $k_t$ declarations preceding $q$ in the same or imported files; 
(ii) \emph{global} sampling of $k_g$ declarations uniformly from $\mathcal{L}$; 
and (iii) \emph{cross-batch} sampling of $k_c$ candidates re-used from other queries in the batch, after masking those that collide with $\mathcal{P}^\star(q)$. The loss is
\[
\mathcal{L}_q \;=\; -\frac{1}{|\mathcal{P}^+_q|} \sum_{p \in \mathcal{P}^+_q}
  \log \frac{\exp\bigl(\mathrm{sim}(h(q),\, h(p))/\tau\bigr)}{\sum_{p' \in \mathcal{P}^+_q \cup \mathcal{P}^-_q}\exp\bigl(\mathrm{sim}(h(q),\, h(p'))/\tau\bigr)},
\]
with temperature $\tau$. Each positive contributes its own log-term but shares the denominator with the rest of $\mathcal{P}^+_q$ and all negatives, encouraging the entire positive group to dominate the ranking rather than only the closest one. 

\subsection{Mathlib-to-Blueprint Adaptation}
\label{sec:adaptation}
\vspace{-5pt}
Lean usage spans diverse distributions (P3). To transfer the structural signal learned on Mathlib4 to the long tail of advanced and not-yet-formalized premises in the Blueprint corpus (Section~\ref{sec:blueprint-mining}), \ourmethod uses a two-stage training procedure. 
\emph{(1) Mathlib4 pretraining} yields strong representations of typed dependency graph structure on the abundant Mathlib data.
\emph{(2) Blueprint continual training} then fine-tunes the structural fusion modules on the Blueprint training split, with a smaller dynamic-sampling budget reflecting the smaller corpus size. 
The queries are informal statements, and the candidate premises are encoded with both their formal and informal statements.
The implementation details are in Appendix~\ref{appx:hyperparams}
We report both the Mathlib-only and the continually-trained model in Section~\ref{sec:experiments}, demonstrating that the typed-graph signal learned on Mathlib transfers, and that Blueprint adaptation further improves recall on rare and advanced premises.

\subsection{Informal-Query Adaptation}
\label{sec:informal}
\vspace{-5pt}
When the query itself is informal (i.e., a natural-language theorem statement), \ourmethod is paired with a formalizer $\mathcal{F}$ that produces a Lean statement $\tilde{q} = \mathcal{F}(q_{\mathrm{nl}})$.
\ourmethod then operates on $\tilde{q}$ as a formal query and returns Lean-side premises. This formalize-then-retrieve composition lets downstream systems that consume informal proofs, such as general-purpose LLMs, still benefit from dependency-aware retrieval, without \ourmethod itself ever processing natural language.

\section{Experiments}
\label{sec:experiments}
\vspace{-5pt}
\subsection{Experimental Setup}
\label{sec:setup}
\vspace{-5pt}
\textbf{Datasets.}
We evaluate on the four benchmarks introduced in Section~\ref{sec:data}: \mlf-Random and \mlf-Heldout for in-distribution and within-Mathlib generalization, \ourmethod-Collection for query-side distribution shift to Olympiad and synthetic problems, and \ourmethod-Blueprint for retrieval against the rare, advanced, and informal+formal premises. 
Split sizes follow Table~\ref{tab:data-splits}.

\textbf{Baselines.}
We group baselines by the experiment in which each is evaluated.
(1)
\textit{Retrieval baselines}: the symbolic selector {MePo}~\citep{meng2009lightweight}, 
the LeanDojo-trained {ReProver}~\citep{yangLeanDojoTheoremProving2023} based on the small variant of Google ByT5~\citep{xue2022byt5}, 
the sentence-transformer-based {LeanHammer}~\citep{zhuLeanHammerPremiseSelection2025}, 
and the general-purpose {Qwen3-Embedding-8B}~\citep{qwen3embedding}. 
For a fair comparison, we retrain ReProver, LeanHammer, and Qwen3-Embedding-8B on our Mathlib dependency data using their published recipes. MePo is symbolic and used as released, via the Lean implementation from
LeanHammer\footnote{\url{https://github.com/hanwenzhu/premise-selection}}.
Every method is run three times with different global random seeds.
ReProver, LeanHammer, and Qwen3-Embedding-8B are deterministic at inference
for a fixed checkpoint, MePo varies with its traversal ordering, and
\ourselector{} varies through dynamic subgraph sampling.
(2)
\textit{Downstream-utility baselines}: in addition to Qwen3-Embedding-8B above, we compare against two production-grade Lean search engines that natively accept informal natural-language queries: {LeanExplore}~\citep{asherLeanExploreSearchEngine2025} and {LeanSearch}~\citep{gaoLeanSearchSemanticSearch2025}. Both are used as released APIs.

\textbf{Metrics.}
For retrieval, we report precision, recall, and macro-F1 at a fixed budget of $K$ retrieved premises. We use $K{=}16$ as the main setting (Table~\ref{tab:main-retrieval}).
Additional results at $K{=}32$ are reported in Appendix~\ref{appx:retrieval-robustness}.
%
For the informal proving experiments, we follow the IMO-ProofBench protocol \citep{imobench} and report grading scores from the IMO-ProofBench autograder, using the same grading prompts and grading model as the original benchmark.

\textbf{Infrastructure.} 
Training and inference compute, as well as hardware specifications, are reported in Appendix~\ref{appx:compute}.

\begin{table*}[t]
  \centering
  \footnotesize
  \setlength{\tabcolsep}{4pt}
  \caption{Retrieval results at $K{=}16$ on all four benchmarks.
    P, R, and F1 denote Precision@16, Recall@16, and macro-F1@16 (\%).
    Values are means over three runs; standard deviations are reported
    in Tables~\ref{tab:full-metrics-mathlib-random}--\ref{tab:full-metrics-blueprint} in Appendix~\ref{appx:retrieval-robustness}.
    Qwen3: Qwen3-Embedding-8B.
    For fair comparison, ReProver, LeanHammer, and Qwen3-Embedding-8B are retrained on the same Mathlib dependency data using their published recipes. MePo is symbolic and used as released. MePo is not applicable to \ourmethod-Blueprint. 
    $\ddagger$: \ourselector{} after Blueprint continual training; all other \ourselector{} results are from the Mathlib-only model.
    Best results are in bold.}
  \label{tab:main-retrieval}
  \begin{tabular}{l rrr rrr rrr rrr}
    \toprule
    & \multicolumn{3}{c}{\textbf{Mathlib4-Random}}
    & \multicolumn{3}{c}{\textbf{Mathlib4-Heldout}}
    & \multicolumn{3}{c}{\textbf{\ourmethod-Collection}}
    & \multicolumn{3}{c}{\textbf{\ourmethod-Blueprint}} \\
    \cmidrule(lr){2-4}\cmidrule(lr){5-7}\cmidrule(lr){8-10}\cmidrule(lr){11-13}
    \textbf{Method} & P & R & F1 & P & R & F1 & P & R & F1 & P & R & F1 \\
    \midrule
    MePo~\citep{mepo}
      & 8.02 & 14.14 & 8.95
      & 4.30 & 7.98 & 5.12
      & 0.74 & 0.53 & 0.60
      & -- & -- & -- \\
    ReProver~\citep{yangLeanDojoTheoremProving2023}
      & 13.01 & 26.97 & 15.68
      & 12.13 & 32.32 & 16.70
      & 1.05 & 0.75 & 0.85
      & 2.44 & 25.37 & 4.36 \\
    LeanHammer~\citep{zhuLeanHammerPremiseSelection2025}
      & 35.56 & 61.13 & 40.57
      & 28.02 & 67.33 & 37.41
      & 2.94 & 1.98 & 2.32
      & 2.97 & 31.36 & 5.32 \\
    Qwen3~\citep{qwen3embedding}
      & 5.84 & 11.54 & 6.87
      & 3.21 & 6.74 & 4.03
      & 0.34 & 0.24 & 0.27
      & 1.11 & 11.02 & 1.97 \\
    \midrule
    \textbf{\ourselector~(Ours)}
      & \textbf{37.60} & \textbf{66.01} & \textbf{43.49}
      & \textbf{32.47} & \textbf{78.74} & \textbf{43.67}
      & \textbf{26.28} & \textbf{13.84} & \textbf{18.00}
      & \textbf{3.83}$^\ddagger$ & \textbf{42.21}$^\ddagger$ & \textbf{6.87}$^\ddagger$ \\
    \bottomrule
  \end{tabular}
\end{table*}

\begin{figure}[t]
  \centering
  \begin{subfigure}[t]{0.32\textwidth}
    \centering
    \includegraphics[width=\linewidth]{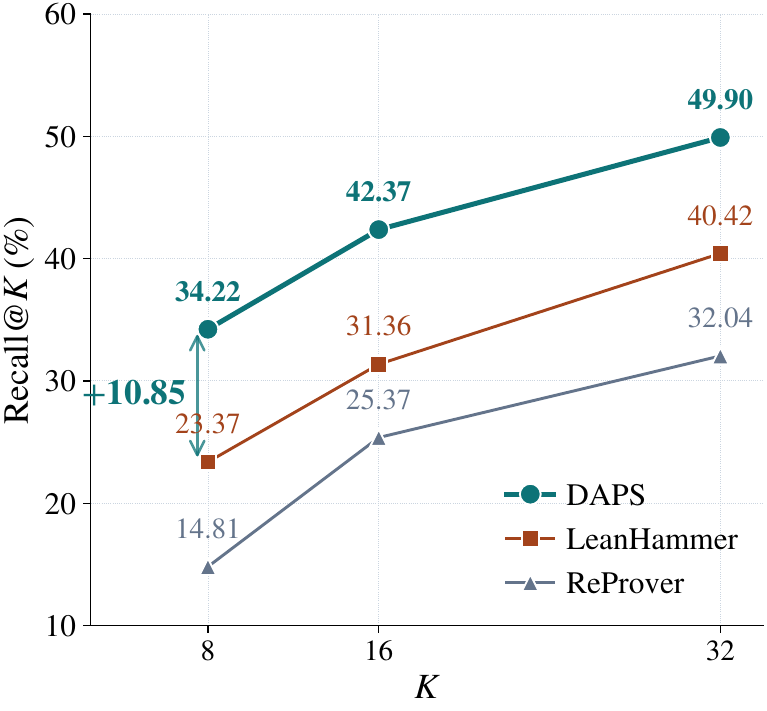}
    \caption{Recall@$K$ on \ourmethod-Blueprint.}
    \label{fig:rk-curve-blueprint}
  \end{subfigure}
  \hfill
  \begin{subfigure}[t]{0.32\textwidth}
    \centering
    \includegraphics[width=\linewidth]{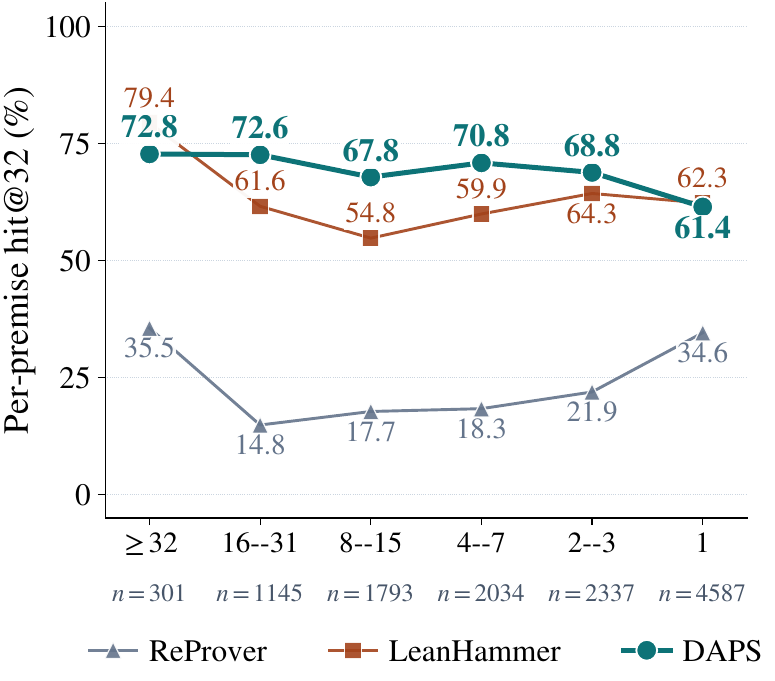}
    \caption{Premise-frequency breakdown on \mlf-Random.}
    \label{fig:premise-frequency-mlf-random}
  \end{subfigure}
  \hfill
  \begin{subfigure}[t]{0.32\textwidth}
    \centering
    \includegraphics[width=\linewidth]{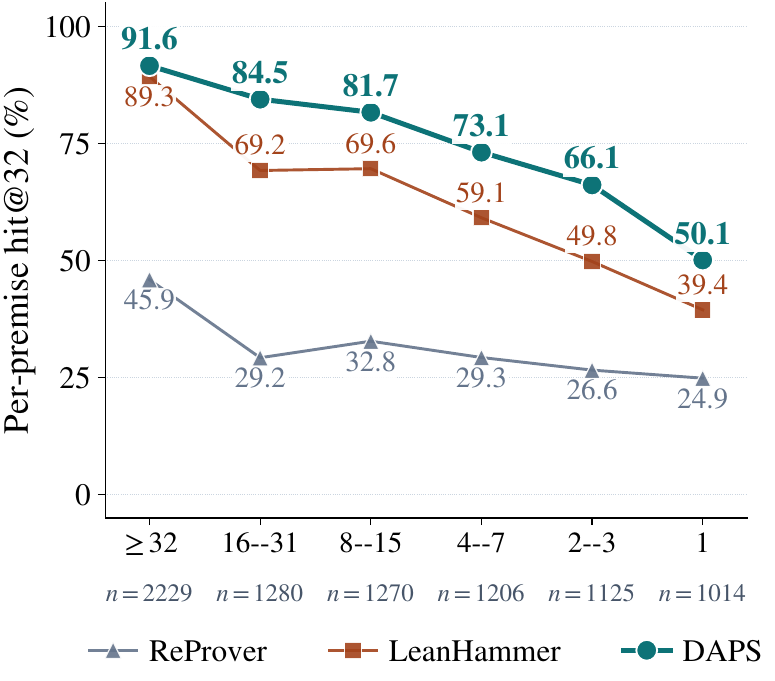}
    \caption{Premise-frequency breakdown on \mlf-Heldout.}
    \label{fig:premise-frequency-mlf-heldout}
  \end{subfigure}
  \caption{
        Single-run comparison of \ourselector{} with ReProver~\cite{yangLeanDojoTheoremProving2023} and LeanHammer~\cite{zhuLeanHammerPremiseSelection2025}.
        (a) Recall@$K$ on \ourmethod-Blueprint for $K \in \{8, 16, 32\}$.
        (b) Per-pair hit@32 by premise frequency on \mlf-Random.
        (c) Per-pair hit@32 by premise frequency on \mlf-Heldout.
    }
  \label{fig:three}
  \vspace{-5pt}
\end{figure}

\begin{table*}[t]
  \small
  \caption{
    Formal theorem proving results on miniF2F-test~\citep{zheng2021minif2f}.
    \textbf{Bold} and \underline{underline} are the best and the second-best results, respectively.
  } 
  \vspace{-10pt}
  \label{table-exp-downstream-formal-proving}
  \begin{center}
    \begin{small}
      \begin{tabular}{l|rrr|rrr}
        \toprule
        & \multicolumn{3}{c|}{\textbf{Goedel-Prover-V2-8B}} & \multicolumn{3}{c}{\textbf{DeepSeek-Prover-V2-7B}} \\
        & pass@1 & pass@2 & pass@4 & pass@1 & pass@2 & pass@4 \\
        \midrule
        w/o premises & 36.48 & 43.62 & 49.90 & 34.07 & 42.23 & 48.23 \\
        LeanExplore~\cite{asherLeanExploreSearchEngine2025}
            & \textbf{40.37} & \underline{47.18} & \underline{52.36} & \underline{36.73} & \underline{44.29} & 48.72 \\
        LeanSearch~\cite{gaoLeanSearchSemanticSearch2025}
            & 39.19 & 46.05 & 51.43 & 34.32 & 42.21 & 47.32 \\
        Qwen3-Embedding-8B~\cite{qwen3embedding}
            & 39.86 & 46.41 & 51.84 & 35.09 & 42.86 & 47.68 \\
        MePo~\cite{mepo} 
            & 37.92 & 44.74 & 50.60 & 35.60 & 43.06 & 47.75 \\
        ReProver~\cite{yangLeanDojoTheoremProving2023}
            & 38.42 & 44.69 & 50.01 & 35.19 & 43.41 & 48.66 \\
        LeanHammer~\cite{zhuLeanHammerPremiseSelection2025}
            & 37.45 & 43.90 & 48.93 & 34.89 & 43.57 & \textbf{49.19} \\
        DAPS & \underline{39.96} & \textbf{47.51} & \textbf{54.03} & \textbf{38.11} & \textbf{44.70} & \underline{49.06} \\
        \bottomrule
      \end{tabular}
    \end{small}
  \end{center}
  \vspace{-5pt}
\end{table*}

\begin{table*}[t]
  \small
  \caption{
    Downstream utility performance.
    Pass@8 of different LLMs on IMO-ProofBench~\citep{imobench} using premises from compared premise selectors. 
    In this table, we strictly follow the original grading prompts from the IMO-ProofBench paper and use the identical grading model.
    Qwen3: Qwen3-Embedding-8B.
    $\Delta$\%: Relative improvement (in \%) of DAPS over the best-performing baseline in each row, computed as (DAPS - best baseline) / best baseline.
    %
    %
  } 
  \vspace{-10pt}
  \label{table-exp-imo-graded-by-gemini}
  \begin{center}
    \begin{small}
      \begin{tabular}{lrrrr|r}
        \toprule
        \textbf{Model} & \textbf{LeanExplore} & \textbf{LeanSearch} & \textbf{Qwen3} & \textbf{\ourselector} & \textbf{$\Delta$\%} \\
        \midrule
        Qwen3-235b~\citep{yangQwen3TechnicalReport2025} & 18.45 & 17.01 & 19.08 & \textbf{21.25} & 11.37 \\
        DeepSeek-V3.1~\citep{deepseek-aiDeepSeekV3TechnicalReport2025} & 11.96 & 12.38 &
        13.75 & \textbf{13.90} & 1.09 \\
        kimi-k2-instruct~\citep{kimik2} & 8.66  & 8.26   & 8.27   & \textbf{9.08} & 4.85 \\
        Gemini-3.6-Flash~\citep{gemini36flash} & 46.13 & 44.04 & 45.51 & \textbf{48.69} & 5.55 \\
        \bottomrule
      \end{tabular}
    \end{small}
  \end{center}
  \vspace{-5pt}
\end{table*}


\subsection{Results on Premise Selection}
\label{sec:main-retrieval}
Table~\ref{tab:main-retrieval} reports precision, recall, and macro-F1 at $K{=}16$ across the four benchmarks. Three observations stand out.
(1) For in-distribution Mathlib retrieval (Mathlib4-Random/Mathlib4-Heldout), \ourselector{} reaches 66.01/78.74 Recall@16, an absolute gain of +4.88/+11.41 over the strong prior selector LeanHammer (61.13/67.33), with corresponding F1 gains of +2.92/+6.26. 
(2) Under out-of-distribution shift to competition-style queries (\ourmethod-Collection), all prior selectors collapse to below 3\% recall, with the best baseline at 1.98. 
\ourselector{} reaches 13.84 recall and 26.28 precision, a $7\times$ and $9\times$ improvement respectively, demonstrating the value of dependency-graph signal under query distribution shift. 
(3) For rare, advanced, and informal+formal premises (\ourmethod-Blueprint), \ourselector{} with Blueprint continual training reaches 42.21 Recall@16, +10.85 over LeanHammer (31.36).

Figure~\ref{fig:rk-curve-blueprint} reports Recall@$K$ on \ourmethod-Blueprint for
$K \in \{8, 16, 32\}$ from a single run, extending the $K{=}16$ comparison of Table~\ref{tab:main-retrieval}. \ourselector{} outperforms LeanHammer at every $K$, with absolute gains of +10.85, +11.01, and +9.48 at $K = 8$, $16$, and $32$, respectively. 
The gap remains large even at small $K$, demonstrating \ourselector{}'s value for downstream LLMs with limited context windows for premises.

\subsection{Results on Formal Theorem Proving}
\label{sec:downstream_formal}
Table~\ref{table-exp-downstream-formal-proving} reports results on miniF2F-test with two whole-proof provers. Implementation details are in Appendix~\ref{appx:impl_formal}.
We report Pass@$k$ for $k \in \{1, 2, 4\}$ because these provers generate a whole proof from a premise list supplied once up front, so a single sample is high-variance.

Supplying \ourselector{}-selected premises improves proof success over the
no-premise baseline in all six settings, with Pass@1 gains of +3.48 on
Goedel-Prover-V2-8B and +4.04 on DeepSeek-Prover-V2-7B. \ourselector{} is
best in four of the six settings and within 0.5 points of the best in the
other two. It also outperforms LeanExplore, a strong baseline, in five of the six
settings. On DeepSeek-Prover-V2-7B the gain narrows as $k$ grows
(+4.04, +2.47, +0.83), and all selectors land within two points of each
other at Pass@4, suggesting that with a larger sampling budget the prover
becomes the bottleneck rather than the premise list.

Other selectors do not consistently help. At Pass@4, LeanHammer falls below
the no-premise baseline with Goedel-Prover-V2-8B (48.93 vs.\ 49.90), as do
LeanSearch, Qwen3-Embedding-8B, and MePo with DeepSeek-Prover-V2-7B. The
same selector can also behave differently across provers. LeanHammer, a strong baseline in retrieval (Table~\ref{tab:main-retrieval}),
gives the best Pass@4 with DeepSeek-Prover-V2-7B but the worst with
Goedel-Prover-V2-8B. End-to-end proof success therefore depends jointly on
the selected premises and the prover's ability to use them, and is not
implied by retrieval metrics alone.
Finally, the evaluated provers were not trained to condition on a supplied premise
list, so these gains are likely a lower bound. Training provers for
premise-conditioned generation is a promising direction we leave to future
work.

\subsection{Results on Informal Theorem Proving}
\label{sec:downstream}
We test whether \ourselector{}'s retrieval improvements also transfer to {informal} proving, where a general-purpose LLM produces a natural-language proof, conditioned on a list of retrieved Lean premises.

\textbf{Setup.}
We follow the adaptation in Section~\ref{sec:informal}. 
Each informal IMO-ProofBench~\citep{imobench} query is first formalized into a Lean statement by Gemini 3 Pro~\citep{gemini3pro}. 
The Lean statement is fed to each selector, which returns its top-16 Mathlib premises. 
The original informal query and the selected premises are then given to a general-purpose LLM, which writes an informal proof. 
The proof is graded by the IMO-ProofBench autograder with the original grading prompt and grading model.
We evaluate four LLMs, Qwen3-235B~\citep{yangQwen3TechnicalReport2025}, DeepSeek-V3.1~\citep{deepseek-aiDeepSeekV3TechnicalReport2025}, Kimi-K2-Instruct~\citep{kimik2}, and Gemini-3.6-Flash~\citep{gemini36flash}. We compare \ourselector{} with LeanExplore~\citep{asherLeanExploreSearchEngine2025}, LeanSearch~\citep{gaoLeanSearchSemanticSearch2025}, and Qwen3-Embedding-8B~\citep{qwen3embedding}.
Every selector receives the same formalized query and the same budget of $K{=}16$ premises.
We report Pass@8 only, owing to the cost of the autograder.



\textbf{Results.}
Table~\ref{table-exp-imo-graded-by-gemini} reports the results.
Premises returned by \ourselector{} yield the highest grades for all four LLMs, with relative gains over the best baseline ranging from 1.09\% (DeepSeek-V3.1) to 11.37\% (Qwen3-235B).
This holds even though the best baseline changes across LLMs (Qwen3-Embedding-8B for Qwen3-235B and DeepSeek-V3.1, LeanExplore for Kimi-K2-Instruct and Gemini-3.6-Flash).
%
The gain on DeepSeek-V3.1 (+0.15 points) is small and may be within grader noise.
%
Still, the consistent ordering across four model families shows that dependency-aware formal retrieval transfers to informal proof generation.

\subsection{Analysis}
\paragraph{Ablation study on training objective.}
To isolate the contribution of the group-level contrastive objective, we replace it with single-positive InfoNCE~\citep{infonce}. This drops Recall@32 to $56.48$ on \mlf-Random, $56.00$ on \mlf-Heldout, and $15.25$ on \ourmethod-Collection, an absolute decrease of $23.14$, $33.30$, and $9.13$ from the full \ourmethod{} configuration. The result confirms the value of group-level relevance as a training signal. Further ablation studies are deferred to Appendix~\ref{appx:ablation}.


\textbf{Premise frequency breakdown.}
On \mlf-Random and \mlf-Heldout, we collect all \texttt{({query} $q$, {gold premise} $p$)} pairs from the test set and bucket each premise by its test-set frequency, i.e., the number of queries that list it as gold. 
For each pair, hit@32 is 1 if $p$ appears in the selector's top-32 retrievals for $q$ and 0 otherwise. We report the mean hit@32 within each bucket on \mlf-Random (Figure~\ref{fig:premise-frequency-mlf-random}) and \mlf-Heldout (Figure~\ref{fig:premise-frequency-mlf-heldout}).
On \mlf-Random, \ourselector and LeanHammer achieve broadly comparable per-bucket hit rates while ReProver trails consistently. LeanHammer narrowly leads on the head bucket and the singletons ($n=1$), while \ourselector leads in the four middle buckets. 
Semantic similarity is therefore sufficient when the training and test premise distributions are closely matched.
The picture changes on \mlf-Heldout, whose file-level isolation introduces a moderate distribution shift on the query side. \ourselector leads LeanHammer in every bucket, by $2.3$ points on the head ($91.6$ vs $89.3$) and $10.7$ on the rarest bucket ($50.1$ vs $39.4$), with the gap widening toward the tail. 
This matches the P1 prediction that rare premises carry less text signal, so baselines miss them more often once the test distribution shifts. 
In contrast, \ourselector lifts performance uniformly across the frequency spectrum.

\textbf{Case studies and limitations.}
We provide additional qualitative case studies and a failure-mode analysis for \ourselector\ in Appendices~\ref{appx:success-cases} and \ref{appx:failure-cases}.
%
Detailed limitations are in Appendix~\ref{appx:limitations}.


\section{Conclusion}
\label{sec:conclusion}
\vspace{-2pt}
We identify three intrinsic limitations of current premise selectors, namely frequency-skewed retrieval (P1), the inability to model group-level relevance (P2), and brittleness under distribution shift (P3). 
All three trace back to underexploiting the dependency structure built into mathematics. 
We curate two underused Lean 4 dependency resources, a typed multi-level graph extracted from Mathlib4 and a systematically mined corpus of Lean Blueprint projects, and derive four retrieval benchmarks of increasing distribution shift. We introduce \ourselector{}, a dependency-aware selector with a structural neighborhood encoder, a group-level contrastive objective, and a Mathlib-then-Blueprint adaptation procedure.
Experiments show that \ourselector{} outperforms strong selectors on all four benchmarks, with +11.41 Recall@16 over LeanHammer on Mathlib4-Heldout and a $7\times$ recall margin on competition-style queries. The lead persists at small retrieval budgets on \ourmethod-Blueprint. 
Downstream, its premises raise Goedel-Prover-V2-8B Pass@4 on miniF2F from 49.90 to 54.03 and improve four general-purpose LLMs on the informal IMO-ProofBench.
These results also suggest that premise-conditioned generation, rather than retrieval alone, is an important remaining bottleneck.

\section*{Acknowledgement and Disclosures}
\vspace{-2pt}
This work was supported in part by the ETH Career Seed Award to Yinya Huang,
the Swiss AI Large Grant SCR1089274,
and Swiss National Supercomputing Centre (CSCS) project ID a130.
Junling Wang is partially supported by an ETH AI Center Doctoral Fellowship.
Yinya Huang is partially supported by an ETH AI Center Postdoctoral Fellowship.

\bibliography{ref_LLMs,ref_ATP,ref_ATP_Benches,ref_ATP_graph,ref_ATP_premise_selection}
\bibliographystyle{plain}


\appendix

\section{Limitations}
\label{appx:limitations}
Our extraction and modeling pipeline is Lean 4 specific. 
Transferring it to Coq, Isabelle, or Agda would require system-specific reimplementation and evaluation. 
Following prior work on premise selection, we use the dependencies of one compiled reference proof as retrieval ground truth.
Theorems that admit alternative proofs may therefore be undercounted by Recall@$K$, which is one reason we also report end-to-end proving results.
Our formal evaluation covers two 7B to 8B whole-proof provers on miniF2F-test. 
It does not cover larger provers, tactic-level or online retrieval, or iterative prover feedback, and the evaluated provers were not trained for premise-conditioned generation. 
Extending \ourselector{} to these settings and to other proof assistants are natural next steps.

\section{Broader Impacts}
\label{appx:broader-impacts}
This work bridges formal and informal mathematical reasoning. 
We mine Lean Blueprints, research-level \LaTeX{} repositories cross-referenced with in-progress Lean formalizations, and use the resulting dependency signal to improve both formal provers (Section~\ref{sec:formal}) and general-purpose LLMs on informal theorem proving (Section~\ref{sec:informal}). 
Positive impacts include narrowing the gap between informal research mathematics and its formal counterpart, and supporting LLM-based mathematical assistants for working mathematicians. 
On the negative side, \ourselector{} reflects the coverage and biases of Mathlib4 and the Blueprint projects we mined. 
Users working in mathematical areas underrepresented in these libraries may see reduced retrieval quality. 
The released artifacts contain only formal and informal mathematical content from public open-source projects, and we do not foresee misuse risks beyond those general to AI systems for mathematical reasoning.

\section{Reproducibility}
\label{appx:reproducibility}
We release the following at the URL\footnote{\url{https://daps-eth.github.io/}} (also presented in the abstract) for review, under a permissive license upon publication:
\begin{itemize}[leftmargin=*, itemsep=2pt]
  \item A demo of the Mathlib4 typed dependency graph and our extraction pipeline.
  \item A demo of the mined Blueprint corpus.
  \item Demos of the four evaluation benchmarks (\mlf-Random, \mlf-Heldout, \ourmethod-Collection, \ourmethod-Blueprint).
  \item the \ourselector{} implementation, training and adaptation scripts, and evaluation pipeline.
\end{itemize}

Training hyperparameters are summarized in Table~\ref{tab:hyperparams}, with optimization details of learning-rate schedule, warmup, batching, and gradient accumulation in Appendix~\ref{appx:hyperparams}.
Compute requirements and runtime statistics for each training stage are reported in Appendix~\ref{appx:compute}. 

\section{Compute Reporting}
\label{appx:compute}
We summarize the compute resources used to produce the experimental results as follows.

\paragraph{Scope.}
This report covers the computation directly used for training the models reported in the main paper. 
The following are explicitly \textbf{not} included: 
(i) the foundation-model pretraining compute of third-party base models we build upon (e.g., MGTE~\citep{zhang2024mgte}); 
(ii) compute used for inference with downstream LLM provers (general-purpose LLMs in Section~\ref{sec:downstream}), which we use as off-the-shelf systems; 
(iii) compute consumed by preliminary experimentation, failed runs, and hyperparameter sweeps not reflected in the final reported results.

\paragraph{Hardware.}
The experimental environment is heterogeneous, comprising a GPU cluster used for the main training stages and a local workstation used for the smaller Blueprint adaptation stage.
\begin{itemize}[leftmargin=1.4em,itemsep=0.1em]
  \item \textbf{Cluster GPUs.} NVIDIA GH200 superchips (Arm Neoverse V2 Grace CPU, 72 CPU cores per superchip; H100 GPU with 96\,GB HBM3 per chip in the configuration we use). Multi-GPU jobs span up to 16 GH200 chips, corresponding to 288 CPU cores per 4-GPU node.
  \item \textbf{Local workstation.} 1$\times$ Apple MacBook Pro (M4 Max, unified-memory architecture).
\end{itemize}

\paragraph{Per-workload compute.}
Table~\ref{tab:compute-summary} summarizes the compute consumed by each training workload. GPU-hours are computed as $(\text{\#GPUs}) \times (\text{wall-clock runtime})$.

\begin{table}[!htbp]
  \centering
  \small
  \caption{Compute consumed by the reported training workloads. Cluster runtimes are aggregated GPU-hours on NVIDIA GH200. The Blueprint adaptation stage is reported as machine-hours since it runs on a single local workstation.}
  \label{tab:compute-summary}
  \setlength{\tabcolsep}{6pt}
  \begin{tabular}{l l r r r}
    \toprule
    \textbf{Workload} & \textbf{Hardware} & \textbf{\#Devices} & \textbf{Runtime} & \textbf{Compute} \\
    \midrule
    Stage 1: text encoder warm-up           & GH200 cluster        & 16   & $\le$12\,h    & $\le$192 GPU-h    \\
    Stage 2: structural fusion training     & GH200 cluster        & 16  & $\le$24\,h   & $\le$384 GPU-h   \\
    Blueprint continual training            & MacBook Pro (M4 Max) & 1   & $\le$3\,h    & $\le$3 machine-h \\
    \midrule
    \textbf{Total (cluster training)}       &                      &     &               & $\le$408 GPU-h   \\
    \textbf{Total (local training)}         &                      &     &               & $\le$3 machine-h \\
    \bottomrule
  \end{tabular}
\end{table}

\paragraph{Notes on the estimates.}
The runtimes above are wall-clock estimates from our training logs and include data-loading, checkpointing, and validation, in addition to optimizer steps. 
They do not include cluster idle/queue time. 
The local Blueprint pipeline takes less than 3 hours end-to-end, of which roughly 2 hours is the training loop itself. 
The remainder is data preprocessing.
All reported hour counts are conservative rounded-up estimates.

\section{Data Schema}
\label{appx:data-schema}
\subsection{Mathlib Graph}
\label{appx:mathlib-graph}
This appendix documents the complete schema of the Mathlib4 dependency graph extracted in Section~\ref{sec:mathlib}. Tables~\ref{table-appx-def-node-cat} and~\ref{table-appx-def-edge-cat} report the per-category counts of all nodes and typed edges in the graph. Tables~\ref{table-appx-def-node-fields} and~\ref{table-appx-def-edge-fields} describe the metadata fields recorded for each node and edge, respectively. Tables~\ref{table-appx-graph-node-indegree} and~\ref{table-appx-graph-note-outdegree} report the in-degree and out-degree distributions broken down by node category.

\paragraph{Edge typing.}
The top 15 typed categories together span 6.5M edges and are dominated by typeclass and instance relations, with \ttt{USES\_CLASS\_DERIVED} and \ttt{USES\_INSTANCE} alone accounting for 1.7M and 1.3M edges, respectively, while proof-level \ttt{DEPENDS\_ON} edges contribute only 618K, less than 10\% of the total. 
Recent neural premise selectors~\citep{yangLeanDojoTheoremProving2023,zhuLeanHammerPremiseSelection2025} typically use these dependency relations as an untyped list of used constants, without distinguishing the dependency categories above. 
\ourselector instead retains edge types as first-class features (Section~\ref{sec:method}).

\begin{figure}[h]
  \centering
  \includegraphics[width=0.50\textwidth]{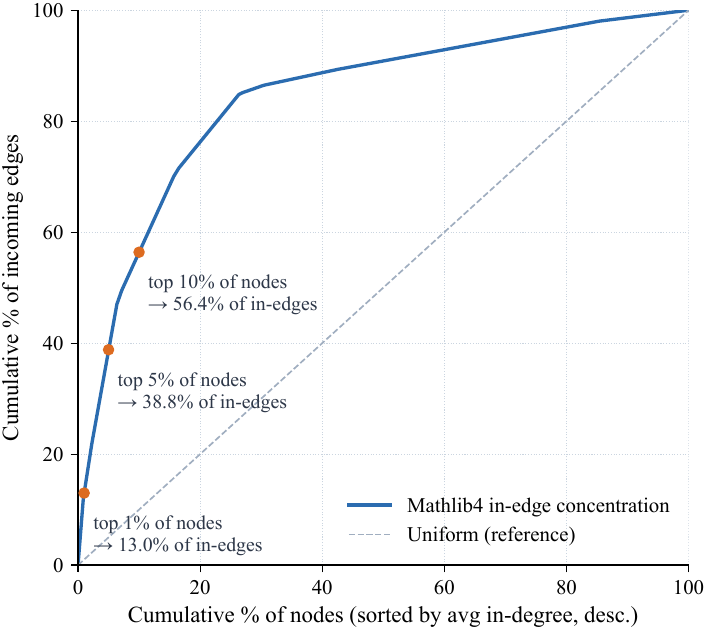}
  \caption{
    The top 10\% of nodes attract 56\% of all incoming edges (category-aggregate approximation; per-node curve is steeper). 
    }
  \vspace{-12pt}
  \label{fig:freqskew}
\end{figure}

\paragraph{Frequency-skewed concentration.}
The asymmetry yields a heavily head-concentrated in-edge distribution (Figure~\ref{fig:freqskew}): the most-cited 1\% of nodes attract 13\% of all incoming dependencies, the top 5\% attract 39\%, and the top 10\% attract 56\%. 
The complement is a low-frequency of theorems and lemmas with very few citations and a correspondingly weak text signal.
This indicates the significance of structural features for accurate retrieval. 

\begin{table}[!htbp]
    \centering
    \begin{minipage}[t]{0.48\textwidth}
        \caption{Statistics of node categories.}
        \label{table-appx-def-node-cat}
          \begin{center}
            \begin{small}
              \begin{tabular}{
                lr
              }
                \toprule
                \textbf{Node Category} & \textbf{Total} \\
        \midrule \midrule
                \ttt{theorem} & 116,096 \\
                \ttt{lemma} & 39,832 \\
                \ttt{definition} & 26,540 \\
                \ttt{instance} & 22,462 \\
                \ttt{definition\_derived} & 15,841 \\
                \ttt{theorem\_derived} & 14,497 \\
                \ttt{class\_derived} & 11,472 \\
                \ttt{structure\_derived} & 9,085 \\
                \ttt{file} & 6,726 \\
                \ttt{lemma\_derived} & 3,918 \\
                \ttt{structure} & 3,260 \\
                \ttt{abbrev} & 2,382 \\
                \ttt{class} & 2,126 \\
                \ttt{instance\_derived} & 1,912 \\
                \ttt{inductive\_derived} & 1,183 \\
                \ttt{others} & 762 \\
                \bottomrule
              \end{tabular}
            \end{small}
          \end{center}        
    \end{minipage}
    \hfill
    \begin{minipage}[t]{0.48\textwidth}
        \caption{Statistics of edge categories.}
          \label{table-appx-def-edge-cat}
          \begin{center}
            \begin{small}
              \begin{tabular}{
                lr
              }
                \toprule
                \textbf{Edge Category} & \textbf{Total} \\
                \midrule \midrule
                \ttt{USES\_CLASS\_DERIVED} & 1,692,183 \\
                \ttt{USES\_INSTANCE} & 1,285,578 \\
                \ttt{USES\_DEFINITION} & 838,828 \\
                \ttt{USES\_CLASS} & 738,150 \\
                \ttt{DEPENDS\_ON} & 618,222 \\
                \ttt{USES\_STRUCTURE} & 567,065 \\
                \ttt{CONTAINS} & 273,116 \\
                \ttt{USES\_INSTANCE\_DERIVED} & 150,501 \\
                \ttt{USES\_ABBREV} & 106,383 \\
                \ttt{USES\_STRUCTURE\_DERIVED} & 77,378 \\
                \ttt{USES\_DEFINITION\_DERIVED} & 75,626 \\
                \ttt{IMPORTS} & 20,794 \\
                \ttt{USES\_INDUCTIVE} & 17,332 \\
                \ttt{USES\_INDUCTIVE\_COMPONENT\_DERIVED} & 15,410 \\
                \ttt{USES\_INDUCTIVE\_DERIVED} & 3,132 \\
             
                \ttt{others} & 3,583 \\
                \bottomrule
              \end{tabular}
            \end{small}
          \end{center}
    \end{minipage}
\end{table}

\begin{table}[!htbp]
  \caption{Description of node fields.
  }
  \label{table-appx-def-node-fields}
  \begin{center}
    \begin{small}
        \begin{tabular}{
          lp{10cm}
        }
          \toprule
          \textbf{Node Field} & \textbf{Definition} \\
          \midrule \midrule
          \texttt{name} & 
            Fully qualified Lean name (e.g., \texttt{AddConstMap.instFunLike}) or the synthetic file\_\* identifiers we create from the filesystem. \\ \midrule
          \texttt{category} &
            Underlying \ttt{ConstantInfo} tag (\ttt{theorem}, \ttt{def}, \ttt{instance}, \ttt{inductive}, \ttt{structure}, \ttt{class}, \ttt{constructor}, \ttt{recursor}, etc.). \\ \midrule
          \texttt{line}/\texttt{column} &
            Start position from \ttt{findDeclarationRanges?}, measured in the source \ttt{.lean} file.\\ \midrule
          \ttt{doc} &
            Docstring preceding the declaration; \ttt{null} when absent. \\ \midrule
          \ttt{signature} & 
            Canonical type with all binders and constants spelled out. \\ \midrule
          \ttt{isProp} & 
            Flags declarations whose type lives in \ttt{Prop}. \\ \midrule
          \ttt{isHumanTheorem} & 
            True only for human-authored theorems that have real source ranges and are not projection functions. \\ \midrule
          \ttt{module} & 
            Module path, e.g., \ttt{Mathlib.Algebra.AddConstMap.Basic}\\ \midrule
          \ttt{source\_path} & 
            Path to the backing \ttt{.lean} file so tools can jump straight into the source tree. \\ \midrule
          \ttt{lean4\_start}/\ttt{lean4\_end} &
            The line where the node's Lean 4 code starts, according to the \ttt{ConstantInfo}. / The line where the node's Lean 4 code ends, decided based on our regex rules.  \\ \midrule
          \ttt{lean4\_code} & The raw Lean 4 text sliced from the \ttt{.lean} between \ttt{lean4\_start} and \ttt{lean4\_end}.\\ \midrule
          \ttt{subcategory} & Enriched category we assigned based on the \ttt{category} and declaration keyword in \ttt{lean4\_code}.   \\ \midrule
          \ttt{statement} & The statement portion of the \ttt{lean4\_code} with comments removed.  \\ \midrule
          \ttt{proof} & Complement to the \ttt{statement}, cpaturing the proof chunk using regex. \\ \midrule
          \ttt{in\_blacklist} & Whether the node is on the curated black list.\\ 
          \bottomrule
        \end{tabular}
    \end{small}
  \end{center}
\end{table}

\begin{table}[!htbp]
  \caption{Description of edge fields.}
  \label{table-appx-def-edge-fields}
  \begin{center}
    \begin{small}
        \begin{tabular}{
          lp{10cm}
        }
          \toprule
          \textbf{Edge Field} & \textbf{Definition} \\
          \midrule \midrule
          \ttt{source\_name}/\ttt{target\_name} & The \ttt{name} entry of source/target node of the edge.\\ \midrule
          \ttt{category} & Semantic label chosen based on the source and the target's \ttt{subcategory}. Conventions: \ttt{CONTAINS} for file $\to$ declaration, \ttt{IMPORTS} for file $\to$ file, \ttt{DEPENDS\_ON} for anything targeting a theorem/lemma/axiom (including \ttt{\_derived} variants), and \ttt{USES\_<CATEGORY>} (e.g., \ttt{USES\_CLASS}, \ttt{USES\_INSTANCE\_DERIVED}, \ttt{USES\_STRUCTURE}) for everything else.
           \\ \midrule
          \ttt{explicit} & Whether a dependency appears in the explicit argument positions of the source declaration. It is \ttt{True} when the referenced constant is used only through explicit parameters( i.e.  \ttt{binderInfo = .default}). It is \ttt{False} when the dependency arises solely via implicit arguments, including implicit \ttt{\{\}} binders or typeclass instance parameters \ttt{[]}. \\ \midrule
          \ttt{direct} & \ttt{True} if the dependency can be obtained from calling function \ttt{getUsedConstantsAsSet} in \ttt{ConstantInfo}). \ttt{False} if it is obtained through further unfolding the direct dependencies of auxiliary lemmas. \\ \midrule
          \ttt{could\_skip} & \ttt{True} if the target node \ttt{in\_blacklist}. \\
          \bottomrule
        \end{tabular}
    \end{small}
  \end{center}
\end{table}

\begin{table}[!htbp]
  \caption{In-degree distribution by node category (sorted by average out-degree).
  }
  \label{table-appx-graph-node-indegree}
  \begin{center}
    \begin{small}
        \begin{tabular}{
          lrrrr
        }
          \toprule
          \textbf{Node Categories} & \textbf{\#node} &	\textbf{avg.} & \textbf{max \#in-degree} & \textbf{total \#in-degree} \\
          \midrule \midrule
          \ttt{class} & 2,126 & 348.20 & 62,237 & 740,276 \\ \midrule
          \ttt{structure} & 3,260 & 174.95 & 24,449 & 570,325 \\ \midrule
          \ttt{class\_derived} & 11,472 & 148.51 & 49,867 & 1,703,655 \\ \midrule
          \ttt{instance\_derived} & 1,912 & 79.71 & 16,527 & 152,413 \\ \midrule
          \ttt{inductive} & 290 & 60.77 & 2,474 & 17,622 \\ \midrule
          \ttt{instance} & 22,462 & 58.24 & 32,159 & 1,308,114 \\ \midrule
          \ttt{abbrev} & 2,382 & 45.66 & 4,768 & 108,765 \\ \midrule
          \ttt{namespace\_command} & 3 & 39.33 & 112 & 118 \\ \midrule
          \ttt{definition} & 26,540 & 32.61 & 58,468 & 865,368 \\ \midrule
          \ttt{section\_command} & 3 & 30.67 & 66 & 92 \\ \midrule
          \ttt{inductive\_component\_derived} & 1,105 & 14.95 & 417 & 16,515 \\ \midrule
          \ttt{abbrev\_derived} & 226 & 13.38 & 382 & 3,025 \\ \midrule
          \ttt{structure\_derived} & 9,085 & 9.52 & 4,472 & 86,463 \\ \midrule
          \ttt{set\_option\_derived} & 10 & 9.00 & 45 & 90 \\ \midrule
          \ttt{definition\_derived} & 15,841 & 5.77 & 5,024 & 91,467 \\ \midrule
          \ttt{theorem\_derived} & 14,497 & 5.75 & 2,388 & 83,368 \\ \midrule
          \ttt{theorem} & 116,096 & 4.90 & 3,106 & 569,172 \\ \midrule
          \ttt{inductive\_derived} & 1,183 & 3.65 & 113 & 4,315 \\ \midrule
          \ttt{namespace\_command\_derived} & 63 & 3.27 & 87 & 206 \\ \midrule          
          \ttt{lemma} & 39,832 & 3.23 & 1,541 & 128,737 \\
          \midrule
          \bottomrule
        \end{tabular}
    \end{small}
  \end{center}
\end{table}

\begin{table}[!htbp]
    \caption{Out-degree distribution by node category (sorted by average out-degree).}
    \label{table-appx-graph-note-outdegree}
    \begin{center}
      \begin{small}
          \begin{tabular}{
            lrrrr
          }
            \toprule
            \textbf{Node Categories} & \textbf{\#node} & \textbf{avg.} & \textbf{max \#out-degree} & \textbf{total \#out-degree} \\
            \midrule \midrule
            \ttt{file} & 6,726 & 43.70 & 684 & 293,910 \\ \midrule
            \ttt{example\_derived} & 1 & 37.00 & 37 & 37 \\ \midrule
            \ttt{lemma} & 39,832 & 31.08 & 395 & 1,237,896 \\ \midrule
            \ttt{theorem} & 116,096 & 26.85 & 474 & 3,116,776 \\ \midrule
            \ttt{lemma\_derived} & 3,918 & 25.36 & 259 & 99,369 \\ \midrule
            \ttt{attribute\_derived} & 3 & 20.33 & 25 & 61 \\ \midrule
            \ttt{constant\_derived} & 2 & 20.00 & 20 & 40 \\ \midrule
            \ttt{theorem\_derived} & 14,497 & 18.66 & 337 & 270,442 \\ \midrule
            \ttt{instance} & 22,462 & 17.94 & 375 & 403,072 \\ \midrule
            \ttt{definition\_derived} & 15,841 & 17.47 & 117 & 276,710 \\ \midrule
            \ttt{definition} & 26,540 & 16.62 & 151 & 441,037 \\ \midrule
            \ttt{abbrev\_derived} & 226 & 14.38 & 62 & 3,251 \\ \midrule
            \ttt{variable\_declaration} & 3 & 14.00 & 15 & 42 \\ \midrule
            \ttt{abbrev} & 2,382 & 13.85 & 91 & 32,996 \\ \midrule
            \ttt{variable\_declaration\_derived} & 94 & 13.67 & 42 & 1,285 \\ \midrule
            \ttt{section\_command} & 3 & 13.00 & 27 & 39 \\ \midrule
            \ttt{structure\_derived} & 9,085 & 12.45 & 135 & 113,129 \\ \midrule
            \ttt{instance\_derived} & 1,912 & 11.94 & 115 & 22,828 \\ \midrule
            \ttt{class\_derived} & 11,472 & 11.02 & 123 & 126,477 \\ \midrule
            \ttt{set\_option\_derived} & 10 & 10.60 & 18 & 106 \\ \midrule
            \ttt{recursor\_derived} & 48 & 10.38 & 23 & 498 \\ \midrule
            \ttt{inductive\_derived} & 1,183 & 8.54 & 74 & 10,098 \\ \midrule
            \ttt{section\_command\_derived} & 37 & 7.32 & 29 & 271 \\ \midrule
            \ttt{namespace\_command\_derived} & 63 & 7.29 & 20 & 459 \\ \midrule
            \ttt{structure} & 3,260 & 5.35 & 84 & 17,444 \\ \midrule
            \ttt{inductive} & 290 & 4.78 & 32 & 1,385 \\ \midrule
            \ttt{inductive\_component\_derived} & 1,105 & 4.71 & 40 & 5,206 \\ \midrule
            \ttt{namespace\_command} & 3 & 4.67 & 9 & 14 \\ \midrule
            \ttt{class} & 2,126 & 3.87 & 50 & 8,231 \\ \midrule
            \ttt{open\_command\_derived} & 34 & 1.24 & 17 & 42 \\ \midrule
            \ttt{initialize} & 38 & 1.05 & 5 & 40 \\ \midrule
            \ttt{constructor} & 2 & 0.50 & 1 & 1 \\ \midrule
            \ttt{opaque} & 107 & 0.45 & 6 & 48 \\ \midrule
            \ttt{opaque\_derived} & 19 & 0.42 & 3 & 8 \\ \midrule
            \ttt{syntax} & 181 & 0.19 & 4 & 35 \\ \midrule
            \ttt{macro\_derived} & 80 & 0.11 & 1 & 9 \\ \midrule
            \ttt{syntax\_derived} & 34 & 0.09 & 1 & 3 \\ \midrule
            \bottomrule
          \end{tabular}
      \end{small}
    \end{center}
  \end{table}

\subsection{Blueprint Mining}
\label{appx-blueprint-list}
\vspace{-10pt}
This appendix lists per-project statistics of all 53 mined Lean Blueprint dependency graphs (Section~\ref{sec:blueprint-mining}), sorted by total node count. Each row corresponds to one blueprint dependency graph.
Project names link to the original blueprint document. The corpus-level summary is reported in Table~\ref{tab:blueprint-summary} of Section~\ref{sec:blueprint-mining}, and the corresponding visualization in Figure~\ref{fig:blueprint-composition}.

\FloatBarrier
{\small
\begin{longtable}{@{}lrrrr@{}}
  \caption{Per-project statistics of the mined Lean Blueprint corpus, sorted by node count. Project names link to the original blueprint.}
  \label{tab:blueprint-per-project}\\
  \toprule
  \textbf{Project} & \textbf{\#Nodes} & \textbf{\#Edges} & \textbf{\#Formal} & \textbf{\#Informal} \\
  \midrule
  \endfirsthead
  \multicolumn{5}{@{}l}{\small\itshape \tablename~\thetable\ continued} \\
  \toprule
  \textbf{Project} & \textbf{\#Nodes} & \textbf{\#Edges} & \textbf{\#Formal} & \textbf{\#Informal} \\
  \midrule
  \endhead
  \midrule
  \multicolumn{5}{r@{}}{\small\itshape continued on next page} \\
  \endfoot
  \bottomrule
  \endlastfoot
  \href{https://AlexKontorovich.github.io/PrimeNumberTheoremAnd/blueprint/dep_graph_document.html}{\texttt{AlexKontorovich/PrimeNumberTheoremAnd}} & 525 & 694 & 487 & 38 \\
  \href{https://RemyDegenne.github.io/brownian-motion/blueprint/dep_graph_document.html}{\texttt{RemyDegenne/brownian-motion}} & 458 & 732 & 298 & 160 \\
  \href{https://teorth.github.io/pfr/blueprint/dep_graph_document.html}{\texttt{teorth/pfr}} & 218 & 403 & 214 & 4 \\
  \href{https://teorth.github.io/equational_theories/blueprint/dep_graph_document.html}{\texttt{teorth/equational\_theories}} & 214 & 185 & 145 & 69 \\
  \href{https://firsching.ch/FormalBook/blueprint/dep_graph_document.html}{\texttt{firsching/FormalBook}} & 189 & 93 & 55 & 134 \\
  \href{https://florisvandoorn.com/carleson/blueprint/dep_graph_document.html}{\texttt{florisvandoorn/carleson}} & 179 & 213 & 175 & 4 \\
  \href{https://Whysoserioushah.github.io/BrauerGroup_new/blueprint/dep_graph_document.html}{\texttt{Whysoserioushah/BrauerGroup\_new}} & 160 & 18 & 119 & 41 \\
  \href{https://YaelDillies.github.io/toric/blueprint/dep_graph_document.html}{\texttt{YaelDillies/toric}} & 145 & 213 & 47 & 98 \\
  \href{https://thefundamentaltheor3m.github.io/Sphere-Packing-Lean/blueprint/dep_graph_document.html}{\texttt{thefundamentaltheor3m/Sphere-Packing-Lean}} & 139 & 201 & 93 & 46 \\
  \href{https://kbuzzard.github.io/ClassFieldTheory/blueprint/dep_graph_document.html}{\texttt{kbuzzard/ClassFieldTheory}} & 120 & 174 & 55 & 65 \\
  \href{https://ilpreterosso.github.io/LEANearized-RadiiPolynomial/blueprint/dep_graph_document.html}{\texttt{ilpreterosso/LEANearized-RadiiPolynomial}} & 99 & 158 & 97 & 2 \\
  \href{https://emilyriehl.github.io/infinity-cosmos/blueprint/dep_graph_document.html}{\texttt{emilyriehl/infinity-cosmos}} & 86 & 104 & 48 & 38 \\
  \href{https://leanprover-community.github.io/liquid/dep_graph_section_1.html}{\texttt{leanprover-community/liquid \S\,s1}} & 77 & 123 & 77 & 0 \\
  \href{https://leanprover-community.github.io/sphere-eversion/blueprint/dep_graph_document.html}{\texttt{leanprover-community/sphere-eversion}} & 73 & 101 & 73 & 0 \\
  \href{https://Ivan-Sergeyev.github.io/seymour/blueprint/dep_graph_document.html}{\texttt{Ivan-Sergeyev/seymour}} & 72 & 137 & 38 & 34 \\
  \href{https://b-mehta.github.io/ABC-Exceptions/blueprint/dep_graph_document.html}{\texttt{b-mehta/ABC-Exceptions}} & 58 & 89 & 35 & 23 \\
  \href{https://mao3j1m0op.github.io/algebra-in-lean/blueprint/dep_graph_document.html}{\texttt{mao3j1m0op/algebra-in-lean}} & 54 & 69 & 54 & 0 \\
  \href{https://sinhp.github.io/HoTTLean/blueprint/dep_graph_document.html}{\texttt{sinhp/HoTTLean}} & 54 & 17 & 17 & 37 \\
  \href{https://b-mehta.github.io/unit-fractions/blueprint/dep_graph_document.html}{\texttt{b-mehta/unit-fractions}} & 52 & 75 & 47 & 5 \\
  \href{https://imperialcollegelondon.github.io/FLT/blueprint/dep_graph_chapter_9.html}{\texttt{imperialcollegelondon/FLT \S\,ch9}} & 52 & 43 & 39 & 13 \\
  \href{https://cameronfreer.github.io/exchangeability/blueprint/dep_graph_document.html}{\texttt{cameronfreer/exchangeability}} & 50 & 69 & 50 & 0 \\
  \href{https://florisvandoorn.com/BonnAnalysis/blueprint/dep_graph_document.html}{\texttt{florisvandoorn/BonnAnalysis}} & 49 & 25 & 14 & 35 \\
  \href{https://YaelDillies.github.io/LeanAPAP/blueprint/dep_graph_document.html}{\texttt{YaelDillies/LeanAPAP}} & 49 & 61 & 37 & 12 \\
  \href{https://ymonbru.github.io/M2/blueprint/dep_graph_document.html}{\texttt{ymonbru/M2}} & 49 & 67 & 2 & 47 \\
  \href{https://leanprover-community.github.io/flt-regular/blueprint/dep_graph_document.html}{\texttt{leanprover-community/flt-regular}} & 45 & 51 & 28 & 17 \\
  \href{https://imperialcollegelondon.github.io/FLT/blueprint/dep_graph_chapter_10.html}{\texttt{imperialcollegelondon/FLT \S\,ch10}} & 39 & 38 & 39 & 0 \\
  \href{https://imperialcollegelondon.github.io/FLT/blueprint/dep_graph_chapter_5.html}{\texttt{imperialcollegelondon/FLT \S\,ch5}} & 35 & 40 & 34 & 1 \\
  \href{https://james18lpc.github.io/GibbsMeasure/blueprint/dep_graph_document.html}{\texttt{james18lpc/GibbsMeasure}} & 35 & 51 & 31 & 4 \\
  \href{https://leanprover-community.github.io/liquid/dep_graph_section_2.html}{\texttt{leanprover-community/liquid \S\,s2}} & 35 & 53 & 31 & 4 \\
  \href{https://vbeffara.github.io/Cours_ED/blueprint/dep_graph_document.html}{\texttt{vbeffara/Cours\_ED}} & 35 & 0 & 1 & 34 \\
  \href{https://AlexandreCharland.github.io/BlueprintRep/blueprint/dep_graph_document.html}{\texttt{AlexandreCharland/BlueprintRep}} & 33 & 42 & 28 & 5 \\
  \href{https://YaelDillies.github.io/ChandraFurstLipton/blueprint/dep_graph_document.html}{\texttt{YaelDillies/ChandraFurstLipton}} & 31 & 49 & 14 & 17 \\
  \href{https://imperialcollegelondon.github.io/FLT/blueprint/dep_graph_chapter_14.html}{\texttt{imperialcollegelondon/FLT \S\,ch14}} & 30 & 22 & 0 & 30 \\
  \href{https://amatelus.com/blueprint/dep_graph_document.html}{\texttt{amatelus/amatelus}} & 27 & 26 & 23 & 4 \\
  \href{https://imperialcollegelondon.github.io/FLT/blueprint/dep_graph_chapter_13.html}{\texttt{imperialcollegelondon/FLT \S\,ch13}} & 21 & 14 & 17 & 4 \\
  \href{https://discreteGeometryInLean.github.io/polytopes-in-LEAN/blueprint/dep_graph_document.html}{\texttt{discreteGeometryInLean/polytopes-in-LEAN}} & 18 & 20 & 11 & 7 \\
  \href{https://e-vergo.github.io/ChebyshevCircles/blueprint/dep_graph_document.html}{\texttt{e-vergo/ChebyshevCircles}} & 18 & 22 & 18 & 0 \\
  \href{https://lean-summer-research.github.io/finalized/blueprint/dep_graph_document.html}{\texttt{lean-summer-research/finalized}} & 17 & 18 & 17 & 0 \\
  \href{https://maybemabeline.github.io/stone-duality/blueprint/dep_graph_document.html}{\texttt{maybemabeline/stone-duality}} & 17 & 2 & 5 & 12 \\
  \href{https://imperialcollegelondon.github.io/FLT/blueprint/dep_graph_chapter_11.html}{\texttt{imperialcollegelondon/FLT \S\,ch11}} & 16 & 21 & 16 & 0 \\
  \href{https://rwst.github.io/infinite-primes-via-log/blueprint/dep_graph_document.html}{\texttt{rwst/infinite-primes-via-log}} & 16 & 13 & 15 & 1 \\
  \href{https://utensil.github.io/lean-ga/blueprint/dep_graph_section_1.html}{\texttt{utensil/lean-ga \S\,s1}} & 16 & 21 & 16 & 0 \\
  \href{https://JonBannon.github.io/Junk-Test/blueprint/dep_graph_document.html}{\texttt{JonBannon/Junk-Test}} & 13 & 8 & 0 & 13 \\
  \href{https://imperialcollegelondon.github.io/FLT/blueprint/dep_graph_chapter_8.html}{\texttt{imperialcollegelondon/FLT \S\,ch8}} & 12 & 13 & 12 & 0 \\
  \href{https://imperialcollegelondon.github.io/FLT/blueprint/dep_graph_chapter_2.html}{\texttt{imperialcollegelondon/FLT \S\,ch2}} & 10 & 11 & 10 & 0 \\
  \href{https://imperialcollegelondon.github.io/FLT/blueprint/dep_graph_chapter_7.html}{\texttt{imperialcollegelondon/FLT \S\,ch7}} & 10 & 7 & 4 & 6 \\
  \href{https://unaoya.github.io/Mills/blueprint/dep_graph_document.html}{\texttt{unaoya/Mills}} & 9 & 5 & 4 & 5 \\
  \href{https://utensil.github.io/lean-ga/blueprint/dep_graph_section_2.html}{\texttt{utensil/lean-ga \S\,s2}} & 9 & 10 & 8 & 1 \\
  \href{https://hanwenzhu.github.io/LeanArchitect-example/blueprint/dep_graph_document.html}{\texttt{hanwenzhu/LeanArchitect-example}} & 8 & 8 & 7 & 1 \\
  \href{https://imperialcollegelondon.github.io/FLT/blueprint/dep_graph_chapter_3.html}{\texttt{imperialcollegelondon/FLT \S\,ch3}} & 8 & 10 & 8 & 0 \\
  \href{https://matijasteblaj.github.io/blossoming-principle/blueprint/dep_graph_document.html}{\texttt{matijasteblaj/blossoming-principle}} & 8 & 0 & 2 & 6 \\
  \href{https://AdamJasonKern.github.io/LBPTester/blueprint/dep_graph_document.html}{\texttt{AdamJasonKern/LBPTester}} & 7 & 7 & 7 & 0 \\
  \href{https://imperialcollegelondon.github.io/FLT/blueprint/dep_graph_chapter_12.html}{\texttt{imperialcollegelondon/FLT \S\,ch12}} & 6 & 5 & 6 & 0 \\
  \midrule
  \textbf{Total (53 projects)} & \textbf{3,805} & \textbf{4,651} & \textbf{2,728} & \textbf{1,077} \\
\end{longtable}}

\subsection{Full Data Splits Method and Evaluation Benchmarks}
\label{appx:data-splits}

\paragraph{Mathlib4 splits.}
The Mathlib dependency graph is divided into 275{,}842 nodes for training, 2{,}000 for validation, and 2{,}000 for testing. The 2{,}000 test nodes consist of two complementary 1{,}000-node slices used as separate benchmarks:
\textbf{\mlf-Heldout} is built by grouping declarations by source file, shuffling the file list, and sequentially selecting files until the cumulative declaration count first exceeds 1{,}000. The corresponding files are excluded from training entirely, and the first 1{,}000 declarations from these files form the test set. This file-level holdout is more conservative than the theorem-level random holdout used by, e.g., LeanHammer~\citep{zhuLeanHammerPremiseSelection2025}, since within-file regularities (shared imports, sibling lemma context) cannot leak through training.
\textbf{\mlf-Random} is built by randomly sampling 1{,}000 declarations from the remaining (non-heldout) files and excluding them from training. 

\paragraph{\ourmethod-Blueprint.}
The Blueprint corpus is split 8:1:1 by node into train/validation/test, after dropping nodes without any recorded dependencies. The resulting test set, \textbf{\ourmethod-Blueprint}, contains 236 queries averaging 1.74 premises each. 
26.0\% are informal-only nodes (\LaTeX\xspace statements without a Lean formalization).
We split by node because the corpus contains only 53 projects with a highly long-tailed size distribution. A project-level split would allocate very few, highly unequal projects to validation and test, making the estimate unstable and strongly dependent on which projects are sampled. \ourmethod-Blueprint therefore measures retrieval of held-out nodes within the mined Blueprint collection, where premises are rare and queries are often informal.


\paragraph{\ourmethod-Collection.}
This dataset is held entirely out of training.
We compile three open-source Lean~4 datasets under \ttt{mathlib-4.22.0} --- 
\texttt{Goedel-LM/Lean-workbook-proofs}\footnote{\url{https://huggingface.co/datasets/Goedel-LM/Lean-workbook-proofs}}~\citep{yingLeanWorkbookLargescale2024,linGoedelProverFrontierModel2025}, 
\texttt{stoney0062/Leanabell-Prover-Traindata-SFT}\footnote{\url{https://huggingface.co/datasets/stoney0062/Leanabell-Prover-Traindata-SFT}}~\citep{zhangLeanabellProverPosttrainingScaling2025}, 
and \texttt{AI-MO/NuminaMath-LEAN}\footnote{\url{https://huggingface.co/datasets/AI-MO/NuminaMath-LEAN}}, yielding 152{,}038 compilable \texttt{(problem, proof)} pairs in Olympiad and synthetic styles. 
We sample 1{,}000 of these and extract their dependency lists from the compiled proofs as ground-truth positives. 
Candidate premises still come from Mathlib, but the query distribution differs substantially from the Mathlib library, giving a clean test of generalization across query distributions while holding the candidate side fixed.

\section{Implementation Details of \ourmethod}
\label{appx:hyperparams}
\paragraph{Mathlib4 pretraining.}
Queries are drawn from the \ttt{theorem} and \ttt{lemma} categories, which are the two declaration types that carry proofs and therefore yield premise lists. For each query, positive premises are taken from its compiled dependencies and prioritized by category in the order
$\ttt{theorem} \succ \ttt{theorem\_derived} \succ \ttt{lemma} \succ \ttt{lemma\_derived} \succ \ttt{definition} \succ \ttt{definition\_derived} \succ \ttt{instance}$;
the top $\ell$ are retained. The semantic encoder reads the \ttt{statement} field for queries, and the concatenation of \ttt{name}, \ttt{category}, \ttt{signature}, \ttt{docstring}, \ttt{statement}, and \ttt{proof} for premises.

Mathlib4 pretraining itself proceeds in two phases. We first warm up the semantic encoder (Section~\ref{sec:struct-encoding}), initialized from MGTE~\citep{zhang2024mgte}, with the masked contrastive loss of \citep{zhuLeanHammerPremiseSelection2025}, contrasting independently-encoded queries and premises. 
We use a cosine learning-rate schedule with warmup ratio $0.04$, per-device train batch size $32$, gradient accumulation steps $2$, and a CachedMNRL mini-batch size of $32$, giving an effective per-step batch of $32 \times 16 \times 2 = 1024$. 

We then freeze the encoder to train the structural fusion with the multi-positive contrastive loss of Section~\ref{sec:group-objective} and all three negative-sampling strategies. 
We use a constant learning-rate schedule with warmup ratio $0.05$, per-device train batch size $16$, and gradient accumulation steps $8$, giving an effective per-step batch of $16 \times 16 \times 8 = 2048$.

Hyperparameters for both phases are listed in Table~\ref{tab:hyperparams}.
Detailed training instances are demonstrated in Figures~\ref{fig:encoder-input-output-ex1} and \ref{fig:rlgt-input-output-ex1}.

\begin{table}[h]
  \centering
  \small
  \caption{Training hyperparameters. $\ell$ = max positives per query (top-$\ell$ after category-priority sort); $m, n$ = dynamic subgraph sampling budget (random sample/keep after FFN scoring); $k_t, k_g, k_c$ = numbers of temporal/global/cross-batch negatives per query.}
  \label{tab:hyperparams}
  \setlength{\tabcolsep}{6pt}
  \begin{tabular}{l rrrrrr}
    \toprule
    & $\ell$ & $m$ & $n$ & $k_t$ & $k_g$ & $k_c$ \\
    \midrule
    Mathlib4 encoder   & 32 & 300 & 64 & 0  & 3  & \texttt{all} \\
    Mathlib4 structural fusion       & 32 & 300 & 64 & 30 & 30 & 1{,}500 \\
    Blueprint continual training      & \texttt{all} & 33 & 33 & 15 & 45 & 960 \\
    \bottomrule
  \end{tabular}
\end{table}

\paragraph{Blueprint continual training.}
The query is encoded from the Blueprint node's informal \LaTeX{} statement only. 
The candidate premise is encoded from its canonical identifier together with any of the following fields that are present: the informal statement, the Lean declaration name, and the formal Lean source.
Reflecting the smaller corpus and lower average out-degree per node, both negative-sampling rates and the dynamic subgraph budget are reduced. 

Continual training runs on a single GPU with a cosine learning-rate schedule (warmup ratio $0.05$), per-device train batch size $16$, and no gradient accumulation, giving an effective per-step batch of $16$. 
Hyperparameters are listed in Table~\ref{tab:hyperparams}. 
Detailed training instance is demonstrated in Figure~\ref{fig:blueprint-input-output-ex1}.




\section{Implementation Details of Formal Theorem Proving with DAPS}
\label{appx:impl_formal}
We conduct end-to-end formal theorem proving in Lean 4 on
miniF2F-test~\citep{zheng2021minif2f}. Besides LeanExplore, LeanSearch, and
Qwen3-Embedding-8B, we also include MePo, ReProver, and LeanHammer as
premise selectors, together with a no-premise setting. For each target
theorem, we use its formal statement as the query and let each selector
retrieve the top-16 premises from Mathlib4. The retrieved premises are
supplied in context to two whole-proof provers,
Goedel-Prover-V2-8B~\citep{linGoedelProverV2ScalingFormal2025} and
DeepSeek-Prover-V2-7B~\citep{deepseekproverv2}. We use temperature 0.7 and
top-$p$ 0.95 for Goedel-Prover-V2-8B, temperature 0.93 and top-$p$ 1.0 for
DeepSeek-Prover-V2-7B, and a maximum of 16,384 output tokens for both. 
The generated proofs are verified with Kimina-Lean-Server~\citep{kimina_lean_server}, and we report Pass@$k$ for $k \in \{1, 2, 4\}$.

Figure~\ref{fig:impl_formal_incontext} shows the prompt used when a
selector supplies 16 premises. The premises are listed by name and
signature and presented as optional hints, so the prover is not required
to use every retrieved item. Figure~\ref{fig:impl_formal_wopremise} shows
the no-premise prompt, which keeps the same formal statement and output
constraint but omits the hint block.

\begin{figure*}[h]
\centering
\begin{tcolorbox}[
  colback=gray!3,
  colframe=black!55,
  boxrule=0.5pt,
  arc=2pt,
  left=6pt,
  right=6pt,
  top=4pt,
  bottom=4pt,
  title={\textbf{Prompt for in-context premise use.}},
  fonttitle=\small,
]

\begin{lstlisting}[breakautoindent=false, breakindent=0pt]
You are an expert in Lean 4 theorem proving. Complete the Lean 4 proof for the statement below.

# Formal statement
```lean4
{formal_statement}
```

# Hints - lemmas that may be related to this goal
These retrieved Lean 4 / Mathlib lemmas may hint at the concepts and dependencies your proof involves. Use them as inspiration rather than a checklist. When one of them is actually the lemma you need, reference it by its full name; otherwise use whatever works (a different lemma or a tactic).
<premise>
[name_1]: [signature_1]
[name_2]: [signature_2]
[name_3]: [signature_3]
...
[name_16]: [signature_16]
</premise>

# Instructions
- Do not use `sorry`. Output only the complete Lean 4 code, no explanation.

# Proof
```lean4
\end{lstlisting}
\end{tcolorbox}
\caption{Prompt for in-context premise use.}
\label{fig:impl_formal_incontext}
\end{figure*}

\begin{figure*}[h]
\centering
\begin{tcolorbox}[
  colback=gray!3,
  colframe=black!55,
  boxrule=0.5pt,
  arc=2pt,
  left=6pt,
  right=6pt,
  top=4pt,
  bottom=4pt,
  title={\textbf{Baseline prompt without retrieved premises.}},
  fonttitle=\small,
]

\begin{lstlisting}[breakautoindent=false, breakindent=0pt]
You are an expert in Lean 4 theorem proving.

# Formal statement
```lean4
{formal_statement}
```

# Task
Write a complete Lean 4 proof for the formal statement. Output only the Lean 4 code.
\end{lstlisting}
\end{tcolorbox}
\caption{Baseline prompt without retrieved premises.}
\label{fig:impl_formal_wopremise}
\end{figure*}

\section{Additional Results}
\subsection{Full Retrieval Results Across Metrics and Seeds}
\label{appx:retrieval-robustness}
Tables~\ref{tab:full-metrics-mathlib-random} to~\ref{tab:full-metrics-blueprint} report precision, recall, and macro-F1 at $K \in \{16, 32\}$ as mean$\pm$standard deviation over three runs with different random seeds. 
A standard deviation of zero indicates deterministic inference for the evaluated checkpoint and query set. 
\ourselector{} ranks first on every reported metric and budget across all four benchmarks.


\subsection{Sensitivity to the Informal-Query Formalizer}
\label{appx:formalizer-ablation}
Our informal proving pipeline (Section~\ref{sec:adaptation}) first
formalizes each natural-language query into a Lean statement before premise
selection. To test how much the downstream result depends on this step, we
vary the formalizer while fixing \ourselector{} as the premise selector and
Gemini-3.6-Flash~\citep{gemini36flash} as the proof generator. We compare
three sources of formal statements. Gemini 3 Pro~\citep{gemini3pro} is the
general-purpose LLM used in Section~\ref{sec:informal}.
Goedel-Formalizer-V2-32B~\citep{linGoedelProverV2ScalingFormal2025} is a formalizer trained specifically for this task. The manual formalizations are the
human-written Lean statements released by Kung et al.\citep{leap}.

Table~\ref{tab:formalizer-ablation} reports Pass@4 on IMO-ProofBench. All three formalizers give comparable results. 
Gemini 3 Pro already serves as a strong formalizer at 46.55, and the specialized and manual formalizations improve this to 47.50 and 48.27.
The downstream benefit of \ourselector{} is therefore not tied to a single formalizer, while lower-noise formal statements bring modest additional gains.



\begin{table*}[h]
  \centering
  \scriptsize
  \setlength{\tabcolsep}{4pt}
  \caption{
  Full retrieval results on Mathlib4-Random, reported as mean $\pm$
  standard deviation over three runs. F1 denotes macro-F1. Best results are
  in bold.
  }
  \label{tab:full-metrics-mathlib-random}
  \resizebox{\textwidth}{!}{%
  \begin{tabular}{lrrrrrr}
    \toprule
    \textbf{Method} & \textbf{Precision@16} & \textbf{Recall@16} & \textbf{F1@16} & \textbf{Precision@32} & \textbf{Recall@32} & \textbf{F1@32} \\
    \midrule
    MePo~\cite{mepo}
        & 8.02$\pm$0.01 & 14.14$\pm$0.03 & 8.95$\pm$0.02 & 5.72$\pm$0.01 & 18.84$\pm$0.04 & 7.93$\pm$0.02 \\
    ReProver~\cite{yangLeanDojoTheoremProving2023} 
        & 13.01$\pm$0.00 & 26.97$\pm$0.00 & 15.68$\pm$0.00 & 9.58$\pm$0.00 & 36.97$\pm$0.00 & 13.99$\pm$0.00 \\
    LeanHammer~\cite{zhuLeanHammerPremiseSelection2025} 
        & 35.56$\pm$0.00 & 61.13$\pm$0.00 & 40.57$\pm$0.00 & 23.45$\pm$0.00 & 73.99$\pm$0.00 & 32.81$\pm$0.00 \\
    Qwen3-Embedding-8B~\cite{qwen3embedding} 
        & 5.84$\pm$0.00 & 11.54$\pm$0.00 & 6.87$\pm$0.00 & 3.91$\pm$0.00 & 15.25$\pm$0.00 & 5.68$\pm$0.00 \\
    \textbf{\ourselector (Ours)} 
        & \textbf{37.60$\pm$0.51} & \textbf{66.01$\pm$0.88} & \textbf{43.49$\pm$0.62} & \textbf{25.71$\pm$0.25} & \textbf{80.68$\pm$0.67} & \textbf{36.01$\pm$0.35} \\
    \bottomrule
  \end{tabular}}
\end{table*}

\begin{table*}[h]
  \centering
  \scriptsize
  \setlength{\tabcolsep}{4pt}
  \caption{
  Full retrieval results on Mathlib4-Heldout, reported as mean $\pm$
  standard deviation over three runs. F1 denotes macro-F1. Best results are
  in bold.
  }
  \label{tab:full-metrics-mathlib-heldout}
  \resizebox{\textwidth}{!}{%
  \begin{tabular}{lrrrrrr}
    \toprule
    \textbf{Method} & \textbf{Precision@16} & \textbf{Recall@16} & \textbf{F1@16} & \textbf{Precision@32} & \textbf{Recall@32} & \textbf{F1@32} \\
    \midrule
    MePo~\cite{mepo} 
        & 4.30$\pm$0.01 & 7.98$\pm$0.01 & 5.12$\pm$0.01 & 2.98$\pm$0.01 & 11.06$\pm$0.01 & 4.37$\pm$0.01 \\
    ReProver~\cite{yangLeanDojoTheoremProving2023}
        & 12.13$\pm$0.00 & 32.32$\pm$0.00 & 16.70$\pm$0.00 & 8.49$\pm$0.00 & 43.48$\pm$0.00 & 13.64$\pm$0.00 \\
    LeanHammer~\cite{zhuLeanHammerPremiseSelection2025}
        & 28.02$\pm$0.00 & 67.33$\pm$0.00 & 37.41$\pm$0.00 & 16.98$\pm$0.00 & 77.84$\pm$0.00 & 26.59$\pm$0.00 \\
    Qwen3-Embedding-8B~\cite{qwen3embedding}
        & 3.21$\pm$0.00 & 6.74$\pm$0.00 & 4.03$\pm$0.00 & 2.34$\pm$0.00 & 9.92$\pm$0.00 & 3.58$\pm$0.00 \\
    \textbf{\ourselector (Ours)} 
    & \textbf{32.47$\pm$0.06} & \textbf{78.74$\pm$0.19} & \textbf{43.67$\pm$0.10} & \textbf{19.67$\pm$0.05} & \textbf{89.25$\pm$0.17} & \textbf{30.79$\pm$0.07} \\
    \bottomrule
  \end{tabular}}
\end{table*}

\begin{table*}[h]
  \centering
  \scriptsize
  \setlength{\tabcolsep}{4pt}
  \caption{
  Full retrieval results on \ourmethod-Collection, reported as mean $\pm$
  standard deviation over three runs. F1 denotes macro-F1. Best results are
  in bold.
  }
  \label{tab:full-metrics-collection}
  \resizebox{\textwidth}{!}{%
  \begin{tabular}{lrrrrrr}
    \toprule
    \textbf{Method} & \textbf{Precision@16} & \textbf{Recall@16} & \textbf{F1@16} & \textbf{Precision@32} & \textbf{Recall@32} & \textbf{F1@32} \\
    \midrule
    MePo~\cite{mepo}
        & 0.74$\pm$0.00 & 0.53$\pm$0.00 & 0.60$\pm$0.00 & 0.53$\pm$0.00 & 0.75$\pm$0.00 & 0.60$\pm$0.00 \\
    ReProver~\cite{yangLeanDojoTheoremProving2023}
        & 1.05$\pm$0.00 & 0.75$\pm$0.00 & 0.85$\pm$0.00 & 0.92$\pm$0.00 & 1.24$\pm$0.00 & 1.03$\pm$0.00 \\
    LeanHammer~\cite{zhuLeanHammerPremiseSelection2025}
        & 2.94$\pm$0.00 & 1.98$\pm$0.00 & 2.32$\pm$0.00 & 2.51$\pm$0.00 & 3.19$\pm$0.00 & 2.76$\pm$0.00 \\
    Qwen3-Embedding-8B~\cite{qwen3embedding}
        & 0.34$\pm$0.00 & 0.24$\pm$0.00 & 0.27$\pm$0.00 & 0.26$\pm$0.00 & 0.36$\pm$0.00 & 0.29$\pm$0.00 \\
    \textbf{\ourselector (Ours)} 
        & \textbf{26.28$\pm$0.48} & \textbf{13.84$\pm$0.24} & \textbf{18.00$\pm$0.32} & \textbf{22.52$\pm$0.73} & \textbf{23.53$\pm$0.76} & \textbf{22.88$\pm$0.74} \\
    \bottomrule
  \end{tabular}}
\end{table*}

\begin{table*}[h]
  \centering
  \scriptsize
  \setlength{\tabcolsep}{4pt}
  \caption{
  Full retrieval results on \ourmethod-Blueprint, reported as mean $\pm$
  standard deviation over three runs. F1 denotes macro-F1. Best results are
  in bold.
  }
  \label{tab:full-metrics-blueprint}
  \resizebox{\textwidth}{!}{%
  \begin{tabular}{lrrrrrr}
    \toprule
    \textbf{Method} & \textbf{Precision@16} & \textbf{Recall@16} & \textbf{F1@16} & \textbf{Precision@32} & \textbf{Recall@32} & \textbf{F1@32} \\
    \midrule
    ReProver~\cite{yangLeanDojoTheoremProving2023}
        & 2.44$\pm$0.00 & 25.37$\pm$0.00 & 4.36$\pm$0.00 & 1.55$\pm$0.00 & 32.04$\pm$0.00 & 2.93$\pm$0.00 \\
    LeanHammer~\cite{zhuLeanHammerPremiseSelection2025}
    & 2.97$\pm$0.00 & 31.36$\pm$0.00 & 5.32$\pm$0.00 & 1.96$\pm$0.00 & 40.42$\pm$0.00 & 3.70$\pm$0.00 \\
    Qwen3-Embedding-8B~\cite{qwen3embedding}
        & 1.11$\pm$0.00 & 11.02$\pm$0.00 & 1.97$\pm$0.00 & 0.65$\pm$0.00 & 13.07$\pm$0.00 & 1.22$\pm$0.00 \\
    \textbf{\ourselector (Ours)} 
        & \textbf{3.83$\pm$0.03} & \textbf{42.21$\pm$0.75} & \textbf{6.87$\pm$0.06} & \textbf{2.42$\pm$0.05} & \textbf{50.08$\pm$0.89} & \textbf{4.55$\pm$0.10} \\
    \bottomrule
  \end{tabular}}
\end{table*}

\begin{table}[h]
  \centering
  \small
  \caption{
  Effect of the informal-query formalizer on IMO-ProofBench. Each
  row changes only the formalizer. The premise selector is \ourselector{}
  and the proof generator is Gemini-3.6-Flash in every row. The metric is
  Pass@4. Best result in \textbf{bold}.
  }
  \label{tab:formalizer-ablation}
  \begin{tabular}{lc}
    \toprule
    \textbf{Query Formalizer} & \textbf{Pass@4} \\
    \midrule
    Gemini-3-Pro & 46.55 \\
    Goedel-Formalizer-V2-32B & 47.50 \\
    Manual formalization & \textbf{48.27} \\
    \bottomrule
  \end{tabular}
\end{table}

\subsection{Ablation Studies}
\label{appx:ablation}

\begin{table*}[t]
  \centering
  \small
  \caption{Component ablation of \ourselector. Recall@32 results are reported. 
  Rows progressively remove components from the full configuration (top row) to isolate each component's contribution. Encoder ablations vary the text encoder backbone. Subgraph ablations vary the dependency-graph signal that \ourselector uses. 
  \textbf{DySS}: Dynamic subgraph sampling.
  \textbf{TLD-only}: Restricted node categories of \texttt{theorem}, \texttt{lemma}, and \texttt{definition}.
  \textbf{Mathlib4-R}: Mathlib4-Random. \textbf{Mathlib4-H}: Mathlib4-Heldout.
  \textbf{\ourmethod-Coll.}: \ourmethod-Collection.
  All rows are single runs with a fixed seed.
  }
  \label{tab:ablation}
  \begin{tabular}{l rrr}
    \toprule
    \textbf{Variant} & \textbf{\mlf-R} & \textbf{\mlf-H} & \textbf{\ourmethod-Coll.} \\
    \midrule
    \multicolumn{4}{l}{\emph{(a) Full configuration}} \\
    \makecell[t]{\ourselector (warmed-up MGTE encoder, DySS, \\ \qquad\qquad\quad 2-hop bidirectional, all node categories)} 
            & \textbf{79.63} & \textbf{89.31} & \textbf{24.38} \\
    \midrule
    \multicolumn{4}{l}{\emph{(b) Encoder ablations}} \\
    \quad $-$ encoder warm-up                                         & 74.30 & 81.08 & 15.63 \\
    \quad $-$ structural fusion (MGTE encoder only)                   & 56.29 & 43.85 &  5.53 \\
    \quad $-$ MGTE encoder $\to$ warmed-up ReProver/ByT5                 & 74.20 & 81.72 &  1.08 \\
    \quad $-$ MGTE encoder $\to$ frozen ReProver/ByT5                  & 59.53 & 64.87 &  4.70 \\
    \midrule
    \multicolumn{4}{l}{\emph{(c) Subgraph design ablations}} \\
    \quad $-$ 2-hop $\to$ 1-hop, out-degree only, with DySS             & 39.81 & 42.14 & 19.84 \\
    \quad $-$ DySS $\to$ Random sampling, 1-hop bidirectional           & 46.74 & 44.49 & 14.99 \\
    \quad $-$ all node categories $\to$ TLD-only, random sampling       & 23.44 & 25.03 &  0.72 \\
    \quad $-$ real graph $\to$ dummy random graph                       & 28.93 & 32.61 &  6.77 \\
    \midrule
    \multicolumn{4}{l}{\emph{(d) Contrastive loss ablation}} \\
    \quad $-$ group-level contrastive objective $\to$ InfoNCE~\citep{infonce}      & 56.48 & 56.00 & 15.25 \\
    \bottomrule
  \end{tabular}
\end{table*}

\label{sec:ablation}
Table~\ref{tab:ablation} reports component ablations of \ourselector{}. We highlight three takeaways.

\paragraph{Structural fusion drives most of the gain.}
Removing the encoder warm-up costs 5 to 9 absolute Recall@32 points across
the three benchmarks, while removing structural fusion entirely costs
23.34/45.46/18.85 points. The gap is largest in relative terms on
\ourmethod-Collection, where structural fusion accounts for 18.85 of the
24.38 total, so the dependency-graph signal, not the encoder backbone, is
what carries retrieval under distribution shift. The backbone still
matters. Replacing MGTE with a ReProver/ByT5 encoder keeps much of the
Mathlib performance but reduces \ourmethod-Collection to 1.08 when warmed
and 4.70 when frozen, indicating that ByT5 is a weaker backbone and that
our warm-up recipe, tuned for MGTE, does not transfer to it.

\paragraph{Multi-hop, bidirectional, all-category subgraphs are needed.}
Each restriction of the subgraph signal substantially degrades
\ourselector{}. Reducing the receptive field to 1-hop out-degree only costs
39.82/47.17 absolute Recall@32 points on the two Mathlib benchmarks.
Replacing dynamic subgraph sampling (DySS) with uniform random sampling
causes further losses, and restricting the sampled neighbors to the
\texttt{theorem}, \texttt{lemma}, and \texttt{definition} categories
degrades all three benchmarks, down to 0.72 on \ourmethod-Collection.
Together, these results show that multi-hop directionality, learned
sampling, and the full typed schema are all necessary for the structural
signal to be useful.


\textbf{The graph signal itself is essential.}
Replacing the real Mathlib dependency graph with a dummy graph (random nodes connected by random edges) collapses \ourselector to 28.93/32.61/6.77 Recall@32 on the two Mathlib benchmarks and \ourmethod-Collection benchmark. 
This demonstrates the value of \ourselector's ability to exploit the typed dependency structure.



\subsection{Successful Retrieval Cases}
\label{appx:success-cases}
This appendix presents five qualitative retrieval case studies covering all four benchmarks of Section~\ref{sec:data}, with two examples from the most diverse benchmark \ourmethod-Blueprint. Each figure shows the query, the gold positive set, the top-10 retrievals from \ourselector\ and the two strongest baselines (LeanHammer and ReProver), and a short commentary. The cases are chosen to illustrate the kinds of structural inferences \ourselector\ makes that text-only baselines miss. Failure cases for \ourselector\ are documented separately in  Appendix~\ref{appx:failure-cases}.

\FloatBarrier
\subsection{Failure Mode Analysis}
\label{appx:failure-cases}
This appendix documents three distinct failure modes of \ourselector\ identified through systematic inspection of low-recall queries across our benchmarks. Table~\ref{tab:failure-modes} summarizes the taxonomy. 
The three modes stem from \ourselector's structural features, with overgeneralization across different dimensions of the dependency graph: type relations, name modifiers, and topical neighborhoods.
We illustrate one representative case per mode below, in the same format as the success cases of Appendix~\ref{appx:success-cases}, with the commentary focused on diagnosing the failure mechanism. 

\begin{table}[h]
  \centering
  \small
  \caption{Failure modes observed in our case studies, with one representative case per mode (full case studies in Figures~\ref{fig:case-study-failure} to \ref{fig:case-study-pnt-failure}).}
  \label{tab:failure-modes}
  \setlength{\tabcolsep}{5pt}
  \begin{tabular}{l p{6.4cm} l}
    \toprule
    \textbf{Failure mode} & \textbf{Symptom} &
    \textbf{Rep. case} \\  
    \midrule
    Type over-generalization
    & \ourselector{} retrieves typeclass instances on a type that
      shares the relevant typeclass with the target, but is not the
      target itself.
    & Figure~\ref{fig:case-study-failure} \\
    \midrule
    Lexical sibling drift
    & When the query is a modifier-prefixed or -suffixed sibling of
      an in-corpus lemma, \ourselector{} retrieves thematically
      related premises rather than the near-identical sibling.
    & Figure~\ref{fig:case-study-blueprint-failure} \\
    \midrule
    Topical subgraph drift
    & \ourselector{} resolves the correct project, but selects the
      wrong topical subgraph when several thematically adjacent
      clusters compete.
    & Figure~\ref{fig:case-study-pnt-failure} \\
    \bottomrule
  \end{tabular}
\end{table}

\FloatBarrier
\begin{figure*}[t]
\centering
\begin{tcolorbox}[
  colback=gray!3,
  colframe=black!55,
  boxrule=0.5pt,
  arc=2pt,
  left=6pt,
  right=6pt,
  top=4pt,
  bottom=4pt,
  title={\textbf{A training instance for semantic encoder warm up}},
  fonttitle=\small,
]

\textbf{Query} \footnotesize\texttt{(statement field of Cardinal.mk\_mul\_le)}
\begin{lstlisting}[basicstyle=\ttfamily\footnotesize,breaklines=true,columns=fullflexible,keepspaces=true,mathescape=true,frame=none,xleftmargin=0pt,aboveskip=2pt,belowskip=2pt]
@[to_additive]
lemma _root_.Cardinal.mk_mul_le : #(s * t) ≤ #s * #t
\end{lstlisting}

\vspace{2pt}\hrule\vspace{4pt}

\textbf{Positive premise} \footnotesize\texttt{(serialized premise text of Set.mul)}
\begin{lstlisting}[basicstyle=\ttfamily\footnotesize,breaklines=true,columns=fullflexible,keepspaces=true,mathescape=true,frame=none,xleftmargin=0pt,aboveskip=2pt,belowskip=2pt]
name: Set.mul

signature: {α : Type u_2} → [Mul α] → Mul (Set α)

lean4_code:
/-- The pointwise multiplication of sets `s * t` and `t` is defined as `{x * y | x ∈ s, y ∈ t}` in
locale `Pointwise`. -/
@[to_additive
      /-- The pointwise addition of sets `s + t` is defined as `{x + y | x ∈ s, y ∈ t}` in locale
      `Pointwise`. -/]
protected def mul : Mul (Set α) :=
  ⟨image2 (· * ·)⟩

scoped[Pointwise] attribute [instance] Set.mul Set.add
\end{lstlisting}

\vspace{2pt}\hrule\vspace{4pt}

\textbf{Sampled negative premises} \footnotesize\texttt{(3 per query, drawn uniformly from the premise pool, with self and known positives filtered out)}
\begin{lstlisting}[basicstyle=\ttfamily\footnotesize,breaklines=true,columns=fullflexible,keepspaces=true,mathescape=true,frame=none,xleftmargin=0pt,aboveskip=2pt,belowskip=2pt]
name: SimpleGraph.Walk.reverse_transfer

signature: ∀ {V : Type u} {G : SimpleGraph V} {u v : V} (p : G.Walk u v) {H : SimpleGraph V} (hp : ∀ (e : Sym2 V), List.instMembership.mem p.edges e → Set.instMembership.mem H.edgeSet e), Eq (p.transfer H hp).reverse (p.reverse.transfer H ⋯)

lean4_code:
@[simp]
... [truncated]

name: CategoryTheory.Limits.PushoutCocone.epi_inr_of_is_pushout_of_epi

signature: ∀ {C : Type u} [inst : CategoryTheory.Category C] {X Y Z : C} {f : inst.Hom X Y} {g : inst.Hom X Z} {t : CategoryTheory.Limits.PushoutCocone f g} (ht : CategoryTheory.Limits.IsColimit t) [CategoryTheory.Epi f], Catego
... [truncated]

name: rank_dual_eq_card_dual_of_aleph0_le_rank'

signature: ∀ {K : Type u} [inst : DivisionRing K] {V : Type u_1} [inst_1 : AddCommGroup V] [inst_2 : Module K V], Cardinal.instLE.le Cardinal.aleph0 (Module.rank K V) → Eq (Module.rank (MulOpposite K) (LinearMap (RingHom.id K) V K)) (Cardinal.mk (Line
... [truncated]
\end{lstlisting}

\end{tcolorbox}
\caption{A training instance for semantic encoder warm up.}
\label{fig:encoder-input-output-ex1}
\end{figure*}

\begin{figure*}[t]
\centering
\begin{tcolorbox}[
  colback=gray!3,
  colframe=black!55,
  boxrule=0.5pt,
  arc=2pt,
  left=6pt,
  right=6pt,
  top=4pt,
  bottom=4pt,
  title={\textbf{A training instance for Mathlib4 pretraining}},
  fonttitle=\small,
]

\textbf{Query} \footnotesize\texttt{(statement field of Cardinal.mk\_mul\_le)}
\begin{lstlisting}[basicstyle=\ttfamily\footnotesize,breaklines=true,columns=fullflexible,keepspaces=true,mathescape=true,frame=none,xleftmargin=0pt,aboveskip=2pt,belowskip=2pt]
@[to_additive]
lemma _root_.Cardinal.mk_mul_le : #(s * t) ≤ #s * #t
\end{lstlisting}

\vspace{2pt}\hrule\vspace{4pt}

\textbf{Positive premise} \footnotesize\texttt{(serialized premise text of Set.image2\_mul)}
\begin{lstlisting}[basicstyle=\ttfamily\footnotesize,breaklines=true,columns=fullflexible,keepspaces=true,mathescape=true,frame=none,xleftmargin=0pt,aboveskip=2pt,belowskip=2pt]
name: Set.image2_mul

category: theorem

signature: ∀ {α : Type u_2} [inst : Mul α] {s t : Set α}, Eq (Set.image2 (fun (x1 x2 : α) => instHMul.hMul x1 x2) s t) (instHMul.hMul s t)

statement: @[to_additive (attr := simp)]
theorem image2_mul : image2 (· * ·) s t = s * t

lean4_code:
@[to_additive (attr := simp)]
theorem image2_mul : image2 (· * ·) s t = s * t :=
  rfl
\end{lstlisting}

\vspace{2pt}\hrule\vspace{4pt}

\textbf{Sampled negative premises} \footnotesize\texttt{(3 shown from sample\_negatives (random\_hard=30, global=30; positives/self filtered))}
\begin{lstlisting}[basicstyle=\ttfamily\footnotesize,breaklines=true,columns=fullflexible,keepspaces=true,mathescape=true,frame=none,xleftmargin=0pt,aboveskip=2pt,belowskip=2pt]
name: Set.natCard_neg

category: lemma_derived

signature: ∀ {G : Type u_1} [inst : InvolutiveNeg G] (s : Set G), Eq (Nat.card (Set.neg.neg s).Elem) (Nat.card s.Elem)

statement: @[to_additive]
lemma natCard_inv (s : Set G) : Nat.card ↥(s⁻¹) = Nat.card s

lean4_code:
@[to_additive]
lemma natCard_inv (s : Set 
... [truncated]

name: Set.ncard_smul_set

category: lemma

signature: ∀ {G : Type u_1} {α : Type u_3} [inst : Group G] [inst_1 : MulAction G α] (a : G) (s : Set α), Eq (instHSMul.hSMul a s).ncard s.ncard

statement: @[to_additive (attr := simp)]
lemma ncard_smul_set (a : G) (s : Set α) : (a • s).ncard = s.ncard

lean4_code:

... [truncated]

name: Set.natCard_smul_set

category: lemma

signature: ∀ {G : Type u_1} {α : Type u_3} [inst : Group G] [inst_1 : MulAction G α] (a : G) (s : Set α), Eq (Nat.card (instHSMul.hSMul a s).Elem) (Nat.card s.Elem)

statement: @[to_additive]
lemma natCard_smul_set (a : G) (s : Set α) : Nat.card ↥(a • s) = Na
... [truncated]
\end{lstlisting}

\end{tcolorbox}
\caption{A training instance for Mathlib4 pretraining.}
\label{fig:rlgt-input-output-ex1}
\end{figure*}

\begin{figure*}[t]
\centering
\begin{tcolorbox}[
  colback=gray!3,
  colframe=black!55,
  boxrule=0.5pt,
  arc=2pt,
  left=6pt,
  right=6pt,
  top=4pt,
  bottom=4pt,
  title={\textbf{A training instance for Blueprint continual training}},
  fonttitle=\small,
]

\textbf{Query} \footnotesize\texttt{(statement field of equational\_theories/partial-1323-sound)}
\begin{lstlisting}[basicstyle=\ttfamily\footnotesize,breaklines=true,columns=fullflexible,keepspaces=true,mathescape=true,frame=none,xleftmargin=0pt,aboveskip=2pt,belowskip=2pt]
Let \({\mathcal F}\) be a partial solution. Then if one defines a partial operation \(\diamond \) on \(M\) by requiring axioms (i), (ii), imposing the additional operations \[ (\phi _a(b)^n x,a) \diamond (\phi _b(c)^n y,b) = (\phi _c(a)^n z,c) \]
\end{lstlisting}

\vspace{2pt}\hrule\vspace{4pt}

\textbf{Positive premise} \footnotesize\texttt{(serialized premise text of equational\_theories/partial-1323)}
\begin{lstlisting}[basicstyle=\ttfamily\footnotesize,breaklines=true,columns=fullflexible,keepspaces=true,mathescape=true,frame=none,xleftmargin=0pt,aboveskip=2pt,belowskip=2pt]
id: partial-1323

nlp: A partial solution is a finite family \({\mathcal F}\) of tuples \((x,y,z,a,b,c) \in (\mathbb {Q}^\times )^3 \times G^6\) with \(a,b,c,0\) distinct, such that the tuples \((\phi _a(b)^n x,a, \phi _b(c)^n y,b)\), \((\phi _b(c)^n y,b,\phi _c(a)^n z,c)\), \((\phi _c(a)^n z,c,\phi _a(b)^{n+1} x, a)\) for \((x,y,z,a,b,c) \in {\mathcal F}\) and \(n \geq 0\) are all distinct.

lean_decl_name: partial-1323
\end{lstlisting}

\vspace{2pt}\hrule\vspace{4pt}

\textbf{Sampled negative premises} \footnotesize\texttt{(3 shown from sample\_negatives (repo=15, global=45; positives/self filtered))}
\begin{lstlisting}[basicstyle=\ttfamily\footnotesize,breaklines=true,columns=fullflexible,keepspaces=true,mathescape=true,frame=none,xleftmargin=0pt,aboveskip=2pt,belowskip=2pt]
id: eq40

nlp: Equation 40 is the law \(0 \diamond 0 \simeq 1 \diamond 1\) (or the equation \(x \diamond x = y \diamond y\)).

lean_decl_name: Equation40

lean4:
def Equation40 (G: Type*) [Magma G] := ∀ x y : G, x ◇ x = 
... [truncated]

id: asterix-obelix-finite

nlp: Any finite magma satisfying the Asterix law (E65) also is left-cancellative and satisfies the Obelix law (E1491).

lean_decl_name: asterix-obelix-finite

id: greedy-iterate

nlp: Every partial solution can be extended to a complete solution that satisfies Equation 2 and Equation 3, and hence 1323.

lean_decl_name: Eq1323.exists_complete_function

lean4:
theorem exists_com
... [truncated]
\end{lstlisting}

\end{tcolorbox}
\caption{A training instance for Blueprint continual training.}
\label{fig:blueprint-input-output-ex1}
\end{figure*}

\FloatBarrier
\begin{figure*}[t]
  \centering
  \begin{tcolorbox}[colback=yellow!2!white, colframe=brown!40, 
                    coltitle=black,
                    title={\textbf{Case study: \texttt{pfr/multidist-ruzsa-III}
                            (\ourselector-Blueprint)}},
                    fontupper=\small]
  
  \textbf{Query (informal statement):}\\
  \textit{Let $m \ge 2$ and let $X_{[m]}$ be a tuple of $G$-valued random 
  variables. If the $X_i$ all have the same distribution, then 
  $D[X_{[m]}] \le m \cdot d[X_i;X_i]$ for any $1 \le i \le m$.}
  
  \medskip
  \textbf{Gold positives (6):}
  \bpref{pfr/ruz-copy}, \bpref{pfr/ruz-indep}, \bpref{pfr/sumset-lower},
  \bpref{pfr/klm-1}, \bpref{pfr/multidist-copy}, \bpref{pfr/multidist-indep}.
  
  \medskip
  
  \begin{minipage}[t]{0.3\textwidth}
    \scriptsize
    \textbf{\ourselector\ top-10:}
    \begin{enumerate}[leftmargin=1.4em, itemsep=0pt, nosep]
      \item \bpref{pfr/ruzsa-diff}
      \item \cmark\ \bpref{pfr/ruz-copy}
      \item \cmark\ \bpref{pfr/ruz-indep}
      \item \bpref{pfr/ruzsa-triangle}
      \item \bpref{pfr/multidist-nonneg}
      \item \bpref{pfr/sym-zero}
      \item \cmark\ \bpref{pfr/sumset-lower}
      \item \bpref{pfr/multidist-def}
      \item \bpref{pfr/ruzsa-nonneg}
      \item \bpref{pfr/multidist-ruzsa-I}
    \end{enumerate}
    \smallskip
    \emph{Recall@32: \textbf{5/6} (0.83)}
  \end{minipage}\hfill
  \begin{minipage}[t]{0.3\textwidth}
    \scriptsize
    \textbf{LeanHammer top-10:}
    \begin{enumerate}[leftmargin=1.4em, itemsep=0pt, nosep]
      \item \bpref{PrimeNumberTheoremAnd/fks2-theorem-3}
      \item \bpref{FormalBook/ch20theorem2}
      \item \bpref{FormalBook/ch16theorem2}
      \item \bpref{pfr/foursum-bound}
      \item \bpref{LeanAPAP/drc}
      \item \bpref{PrimeNumberTheoremAnd/bklnw-cor-3-1}
      \item \bpref{pfr/phi-first-estimate}
      \item \bpref{PrimeNumberTheoremAnd/buthe-theorem-2f}
      \item \bpref{PrimeNumberTheoremAnd/buthe-theorem-2d}
      \item \bpref{pfr/ruzsa-diff}
    \end{enumerate}
    \smallskip
    \emph{Recall@32: \textbf{0/6} (0.00)}
  \end{minipage}\hfill
  \begin{minipage}[t]{0.3\textwidth}
    \scriptsize
    \textbf{ReProver top-10:}
    \begin{enumerate}[leftmargin=1.4em, itemsep=0pt, nosep]
      \item \bpref{FormalBook/ch38lemma2}
      \item \bpref{carleson/antichain-decomposition}
      \item \bpref{pfr/gen-ineq}
      \item \bpref{ABC-Exceptions/prop:ThueEquations}
      \item \bpref{toric/1-1-phiA}
      \item \bpref{polytopes-in-LEAN/prop:v_is_h}
      \item \bpref{PrimeNumberTheoremAnd/Dedekind-nonvanishing}
      \item \bpref{FormalBook/ch22lemma1}
      \item \bpref{PrimeNumberTheoremAnd/Dedekind-PNT}
      \item \bpref{equational_theories/axiom-b}
    \end{enumerate}
    \smallskip
    \emph{Recall@32: \textbf{0/6} (0.00)}
  \end{minipage}
  
  \medskip
  \textbf{Commentary.}
  \ourselector\ retrieves \bpref{ruz-copy} and \bpref{ruz-indep}, the two 
  defining lemmas linking Ruzsa distance $d[\cdot;\cdot]$ to entropies, in positions 2 and 3, and surfaces 5 of the 6 gold premises within its top-16. LeanHammer and ReProver miss \emph{all} 6 gold premises within their top-32, instead of retrieving lemmas from unrelated Blueprint projects 
  (\bpref{FormalBook}, \bpref{PrimeNumberTheoremAnd}, \bpref{carleson}, 
  \bpref{ABC-Exceptions}, \bpref{toric}). The gold premises form a tight subgraph in PFR's Ruzsa-distance section. \ourselector's typed dependency edges keep retrieval inside this neighborhood, while the text-only baselines are pulled across the corpus by surface similarity to other entropy and inequality lemmas.
  
  \end{tcolorbox}
  \caption{Case study from the \texttt{teorth/pfr} Blueprint project. 
  \cmark\ marks premises in the gold positive set. 
  Premise names follow the PFR Blueprint label scheme.
  }
  \label{fig:case-study-pfr}
\end{figure*}

\begin{figure*}[t]
  \centering
  \begin{tcolorbox}[
    colback=yellow!2!white, colframe=brown!40, 
    coltitle=black,
    title={\textbf{Case study: \texttt{PrimeNumberTheoremAnd/fks-corollary-3-12} (\ourselector-Blueprint)}},
    fontupper=\small
  ]
  
  \textbf{Query (informal statement):}\\
  \textit{Let \(5/8 < \sigma _2 \leq 1\), \(t_0 = t_0(\sigma _2, x) = \max \left(H_{\sigma _2}, \exp \left(\sqrt{\frac{\log x}{R}}\right)\right)\), \(T > t_0\). Let \(K \geq 2\), \(\lambda = (T/t_0)^{1/K}\), and consider \((t_k)_{k=0}^K\) the sequence given by \(t_k = t_0 \lambda ^k\). Then \[ \Sigma ^1_{\sigma _2} = 2 \sum _{\substack {0 < \gamma < T \\ \sigma _2 \leq \beta < 1}} \frac{x^{\beta -1}}{\gamma } \leq \varepsilon _4(x, \sigma _2, K, T), \]}
  
  \medskip
  \textbf{Gold positives (4):}
  \bpref{PrimeNumberTheoremAnd/zero-density-bound}, 
  \bpref{PrimeNumberTheoremAnd/Riemann-von-Mangoldt-estimate},
  \bpref{PrimeNumberTheoremAnd/RH-up-to}, 
  \bpref{PrimeNumberTheoremAnd/classical-zero-free-region}.
  
  \medskip
  
  \begin{minipage}[t]{0.3\textwidth}
    \scriptsize
    \textbf{\ourselector\ top-10:}
    \begin{enumerate}[leftmargin=1.4em, itemsep=0pt, nosep]
      \item \cmark\ \bpref{PrimeNumberTheoremAnd/zero-density-bound}
      \item \cmark\ \bpref{PrimeNumberTheoremAnd/Riemann-von-Mangoldt-estimate}
      \item \cmark\ \bpref{PrimeNumberTheoremAnd/RH-up-to}
      \item \bpref{PrimeNumberTheoremAnd/Etheta-def}
      \item \bpref{PrimeNumberTheoremAnd/bklnw-eq_A_16}
      \item \bpref{ABC-Exceptions/lem:SubSumGeometry}
      \item \bpref{PrimeNumberTheoremAnd/Epsi-def}
      \item \bpref{ABC-Exceptions/lem:Trivialaibici}
      \item \bpref{PrimeNumberTheoremAnd/classical-bound-psi}
      \item \bpref{Sphere-Packing-Lean/thm:ramanujan-formula}
    \end{enumerate}
    \smallskip
    \emph{Recall@32: \textbf{4/4} (1.00)}
  \end{minipage}\hfill
  \begin{minipage}[t]{0.3\textwidth}
    \scriptsize
    \textbf{LeanHammer top-10:}
    \begin{enumerate}[leftmargin=1.4em, itemsep=0pt, nosep]
      \item \bpref{PrimeNumberTheoremAnd/fks-proposition-3-11}
      \item \bpref{PrimeNumberTheoremAnd/fks-proposition-3-14}
      \item \bpref{PrimeNumberTheoremAnd/fks2-lemma-10b}
      \item \bpref{PrimeNumberTheoremAnd/bklnw-eq_A_13}
      \item \bpref{PrimeNumberTheoremAnd/bklnw-eq_A_10}
      \item \bpref{PrimeNumberTheoremAnd/fks2-corollary-26}
      \item \bpref{PrimeNumberTheoremAnd/bklnw-cor-3-1}
      \item \bpref{carleson/Holder-van-der-Corput}
      \item \bpref{PrimeNumberTheoremAnd/bklnw-eq_A_12}
      \item \bpref{PrimeNumberTheoremAnd/buthe-theorem-2d}
    \end{enumerate}
    \smallskip
    \emph{Recall@32: \textbf{0/4} (0.00)}
  \end{minipage}\hfill
  \begin{minipage}[t]{0.3\textwidth}
    \scriptsize
    \textbf{ReProver top-10:}
    \begin{enumerate}[leftmargin=1.4em, itemsep=0pt, nosep]
      \item \bpref{toric/1-2-14-dual-rat-cone}
      \item \bpref{ABC-Exceptions/def:nu}
      \item \bpref{PrimeNumberTheoremAnd/lem:abadimpseri}
      \item \bpref{PrimeNumberTheoremAnd/bklnw-thm_16}
      \item \bpref{PrimeNumberTheoremAnd/lem:aachcanc}
      \item \bpref{FormalBook/ch26lemma_d}
      \item \bpref{PrimeNumberTheoremAnd/bklnw-prop-3-sub-1}
      \item \bpref{PrimeNumberTheoremAnd/bklnw-cor-5-1}
      \item \bpref{PrimeNumberTheoremAnd/lem:abadeulmac'}
      \item \bpref{PrimeNumberTheoremAnd/symm_inv_log}
    \end{enumerate}
    \smallskip
    \emph{Recall@32: \textbf{0/4} (0.00)}
  \end{minipage}
  
  \medskip
  \textbf{Commentary.}
  \ourselector\ retrieves \emph{all four} gold premises in its top-32 (positions 1, 2, 3, and 16), giving perfect recall. The two failure modes of the baselines are revealing. 
  LeanHammer stays inside \bpref{PrimeNumberTheoremAnd} but is captured by surface similarity to the query name: it returns \bpref{fks-proposition-3-11/14} and several \bpref{bklnw-eq_*} entries from neighboring sections of the same paper, missing the gold premises that live in the earlier zero-density and zero-free-region sections. ReProver drifts further, retrieving lemmas from \bpref{toric} and \bpref{ABC-Exceptions} entirely outside the relevant subgraph. 
  \ourselector's typed dependency edges resolve the precise mathematical neighborhood of zero-density estimates and zero-free regions of $\zeta$, that this corollary actually consumes.
  
  \end{tcolorbox}
  \caption{Case study from the \texttt{AlexKontorovich/PrimeNumberTheoremAnd} 
  Blueprint project. \cmark\ marks premises in the gold positive set. Premise names follow the PFR Blueprint label scheme.}
  \label{fig:case-study-pnt}
\end{figure*}
\begin{figure*}[t]
  \centering
  \begin{tcolorbox}[
    colback=yellow!2!white, colframe=brown!40, 
    coltitle=black,
    title={\textbf{Case study: \texttt{lean\_workbook\_plus\_39047} 
           (cosine difference formula, \ourmethod-Collection)}},
    fontupper=\small
  ]
  
  \textbf{Query (Lean 4 statement):}
  \begin{lstlisting}[basicstyle=\ttfamily\footnotesize]
theorem lean_workbook_plus_39047 (x y : R) :
    cos (x - y) = cos x * cos y + sin x * sin y := by
  \end{lstlisting}
  
  \medskip
  \textbf{Gold positives (16, ranked by priority):}
  \bpref{Real.cos}, \bpref{Real.instSub}, \bpref{Real.instAdd}, 
  \bpref{Real.cos_sub}, \bpref{Real.sin}, \bpref{AddGroup}, 
  \bpref{neg_add_cancel}, \bpref{add_neg_cancel}, \bpref{lt_trans}, 
  \bpref{IsStrictOrderedRing}, \dots\ (16 total).
  
  \medskip
  
  \begin{minipage}[t]{0.3\textwidth}
    \scriptsize
    \textbf{\ourselector\ top-10:}
    \begin{enumerate}[leftmargin=1.4em, itemsep=0pt, nosep]
      \item \bpref{Real.instRatCast}
      \item \cmark\ \bpref{Real.instSub}
      \item \bpref{Real.pseudoMetricSpace}
      \item \bpref{Real.instMul}
      \item \cmark\ \bpref{Real.instAdd}
      \item \bpref{Real.instRing}
      \item \bpref{Real.instNeg}
      \item \bpref{Real.instDivisionRing}
      \item \bpref{Real.instSemigroup}
      \item \bpref{Real.normedCommRing}
    \end{enumerate}
    \smallskip
    \emph{Recall@32: \textbf{4/16} (0.25)}
  \end{minipage}\hfill
  \begin{minipage}[t]{0.3\textwidth}
    \scriptsize
    \textbf{LeanHammer top-10:}
    \begin{enumerate}[leftmargin=1.4em, itemsep=0pt, nosep]
      \item \bpref{Complex.cos}
      \item \cmark\ \bpref{Real.cos}
      \item \bpref{Complex.sin}
      \item \bpref{Complex.cos_neg}
      \item \cmark\ \bpref{Real.sin}
      \item \bpref{Complex.cos_add}
      \item \bpref{Complex.ofReal_cos}
      \item \bpref{Complex.ofReal_sin}
      \item \bpref{Complex.sin_add}
      \item \bpref{Real.cos_add}
    \end{enumerate}
    \smallskip
    \emph{Recall@32: \textbf{3/16} (0.19)}
  \end{minipage}\hfill
  \begin{minipage}[t]{0.3\textwidth}
    \scriptsize
    \textbf{ReProver top-10:}
    \begin{enumerate}[leftmargin=1.4em, itemsep=0pt, nosep]
      \item \cmark\ \bpref{Real.cos}
      \item \bpref{Complex.ofReal_cos}
      \item \bpref{Complex.sin}
      \item \bpref{Complex.cos}
      \item \bpref{Complex.ofReal_sin}
      \item \bpref{Complex.sin_eq_zero_iff_cos_eq}
      \item \bpref{Complex.cos_eq_zero_iff_sin_eq}
      \item \bpref{Complex.sin_ofReal_re}
      \item \bpref{Real.cos_zero}
      \item \cmark\ \bpref{Real.sin}
    \end{enumerate}
    \smallskip
    \emph{Recall@32: \textbf{2/16} (0.12)}
  \end{minipage}
  
  \medskip
  \textbf{Commentary.}
  All three selectors hit at least one gold premise here, but the top-10 reveals different retrieval behaviors. \ourselector's top-10 is dominated by \bpref{Real.inst*} typeclass instances (8 of 10 entries), which is the algebraic infrastructure on which the cosine-difference proof actually rewrites. LeanHammer surfaces both \bpref{Real.cos} and \bpref{Real.sin} in its top-5 but wastes the remaining slots on \bpref{Complex.*} duplicates (6 of 10 entries) of the same trigonometric lemmas. ReProver shows the same \bpref{Complex.*} drift more severely. The two signals are complementary: textual encoders catch the namesake trig lemmas faster, but \ourselector's typed dependency edges surface the typeclass and structural premises that proofs of competition-style theorems also depend on, yielding the highest overall Recall@32.
  
  \end{tcolorbox}
  \caption{Case study from the \ourmethod-Collection benchmark 
(a Lean Workbook problem). \cmark\ marks premises in the gold positive 
set. Premise names follow Mathlib4 declaration paths.}
  \label{fig:case-study-collection}
\end{figure*}
\begin{figure*}[t]
  \centering
  \begin{tcolorbox}[
    colback=yellow!2!white, colframe=brown!40, 
    coltitle=black,
    title={\textbf{Case study: \texttt{add\_lt\_add\_of\_lt\_of\_lt} 
           (\mlf-Heldout)}},
    fontupper=\small
  ]
  
  \textbf{Query (Lean 4 source, additive version generated by 
  \texttt{@[to\_additive]}):}
  \begin{lstlisting}[basicstyle=\ttfamily\footnotesize]
@[to_additive (attr := gcongr high)]
theorem mul_lt_mul_of_lt_of_lt [MulLeftStrictMono α]
    [MulRightStrictMono α]
    {a b c d : α} (h₁ : a < b) (h₂ : c < d) :
    a * c < b * d
\end{lstlisting}
  
  \medskip
  \textbf{Gold positives (6):}
  \bpref{add_lt_add_left}, \bpref{add_lt_add_right}, 
  \bpref{instTransLt_mathlib}, \bpref{AddLeftStrictMono}, 
  \bpref{AddRightStrictMono}, \bpref{Preorder}.
  
  \medskip
  
  \begin{minipage}[t]{0.3\textwidth}
    \scriptsize
    \textbf{\ourselector\ top-10:}
    \begin{enumerate}[leftmargin=1.4em, itemsep=0pt, nosep]
      \item \cmark\ \bpref{Preorder}
      \item \bpref{MulLeftStrictMono}
      \item \bpref{MulRightStrictMono}
      \item \cmark\ \bpref{AddRightStrictMono}
      \item \cmark\ \bpref{AddLeftStrictMono}
      \item \bpref{mul_lt_mul_of_lt_of_lt}
      \item \cmark\ \bpref{add_lt_add_left}
      \item \cmark\ \bpref{add_lt_add_right}
      \item \bpref{mul_lt_mul_left'}
      \item \bpref{mul_lt_mul_right'}
    \end{enumerate}
    \smallskip
    \emph{Recall@32: \textbf{6/6} (1.00)}
  \end{minipage}\hfill
  \begin{minipage}[t]{0.3\textwidth}
    \scriptsize
    \textbf{LeanHammer top-10:}
    \begin{enumerate}[leftmargin=1.4em, itemsep=0pt, nosep]
      \item \bpref{gauge}
      \item \bpref{gauge_def}
      \item \bpref{gauge_def'}
      \item \bpref{Set.instMembership}
      \item \bpref{Module}
      \item \bpref{gauge_nonneg}
      \item \bpref{Real.instInfSet}
      \item \bpref{gauge_zero}
      \item \bpref{Set}
      \item \bpref{InfSet.sInf}
    \end{enumerate}
    \smallskip
    \emph{Recall@32: \textbf{0/6} (0.00)}
  \end{minipage}\hfill
  \begin{minipage}[t]{0.3\textwidth}
    \scriptsize
    \textbf{ReProver top-10:}
    \begin{enumerate}[leftmargin=1.4em, itemsep=0pt, nosep]
      \item \bpref{min_lt_max_of_mul_lt_mul}
      \item \bpref{Left.min_le_max_of_mul_le_mul}
      \item \bpref{Right.min_le_max_of_mul_le_mul}
      \item \bpref{LinearOrder.lift}
      \item \bpref{max_mul_min}
      \item \bpref{max_min_distrib_right}
      \item \bpref{max_min_distrib_left}
      \item \bpref{min_le_max_of_mul_le_mul}
      \item \bpref{fn_max_mul_fn_min}
      \item \bpref{min_mul_max}
    \end{enumerate}
    \smallskip
    \emph{Recall@32: \textbf{0/6} (0.00)}
  \end{minipage}
  
  \medskip
  \textbf{Commentary.}
  \ourselector\ retrieves all six gold premises within its top-11 and five 
  of them within its top-10. The remaining four entries in \ourselector's 
  top-10 are the multiplicative siblings 
  (\bpref{MulLeftStrictMono}, \bpref{MulRightStrictMono}, 
  \bpref{mul_lt_mul_of_lt_of_lt}, \bpref{mul_lt_mul_left'}, 
  \bpref{mul_lt_mul_right'}), which are the \texttt{@[to\_additive]} 
  source declarations from which the additive query is generated. 
  LeanHammer drifts to the unrelated \bpref{gauge} namespace (a functional 
  analysis context) and loses the strict monotonicity meaning entirely. 
  ReProver stays closer in spirit by retrieving \bpref{min}/\bpref{max} 
  lemmas about multiplication, but still misses all six gold premises. 
  The dependency graph captures the \texttt{@[to\_additive]} inheritance 
  link and lifts both the multiplicative parent and the additive children 
  into top positions, while text-only retrieval reaches for surface 
  similarity and ends up in adjacent but unrelated parts of Mathlib.
  
  \end{tcolorbox}
  \caption{Case study from the \mlf-Held-out benchmark. \cmark\ marks 
  premises in the gold positive set. Premise names are Mathlib4 declaration 
  paths.}
  \label{fig:case-study-mathlib}
\end{figure*}
\begin{figure*}[t]
  \centering
  \begin{tcolorbox}[
    colback=yellow!2!white, colframe=brown!40, 
    coltitle=black,
    title={\textbf{Case study: \texttt{Algebra.map\_leftMulMatrix\_localization} 
           (\mlf-Random)}},
    fontupper=\small
  ]
  
  \textbf{Query (Lean 4 statement):}
  \begin{lstlisting}[basicstyle=\ttfamily\footnotesize]
theorem Algebra.map_leftMulMatrix_localization 
    {ι : Type*} [Fintype ι] [DecidableEq ι]
    (b : Basis ι R S) (a : S) :
    (algebraMap R Rₘ).mapMatrix (leftMulMatrix b a) =
    leftMulMatrix (b.localizationLocalization Rₘ M Sₘ) 
                  (algebraMap S Sₘ a)
  \end{lstlisting}
  
  \medskip
  \textbf{Gold positives (32, ranked by priority):}
  \bpref{Algebra.leftMulMatrix_eq_repr_mul}, \bpref{Matrix.ext}, 
  \bpref{Module.Basis.localizationLocalization_apply}, 
  \bpref{Module.Basis.localizationLocalization_repr_algebraMap}, 
  \bpref{Algebra.algebraMapSubmonoid}, \bpref{Algebra.leftMulMatrix}, 
  \bpref{Matrix}, \bpref{Module.Basis.localizationLocalization}, 
  \bpref{RingHom.mapMatrix}, \bpref{algebraMap}, \dots\ (32 total).
  
  \medskip
  
  \begin{minipage}[t]{0.3\textwidth}
    \scriptsize
    \textbf{\ourselector\ top-10:}
    \begin{enumerate}[leftmargin=1.4em, itemsep=0pt, nosep]
      \item \cmark\ \bpref{Algebra.algebraMapSubmonoid}
      \item \cmark\ \bpref{Module.Basis.localizationLocalization}
      \item \bpref{IsScalarTower.algebraMap_apply}
      \item \cmark\ \bpref{Module.Basis}
      \item \bpref{Module.Basis.coe_mapCoeffs}
      \item \bpref{RingEquiv.ofBijective}
      \item \cmark\ \bpref{Module.Basis.instFunLike}
      \item \cmark\ \bpref{Algebra.leftMulMatrix}
      \item \bpref{Algebra.smul_def}
      \item \bpref{Algebra.toModule}
    \end{enumerate}
    \smallskip
    \emph{Recall@32: \textbf{11/32} (0.34)}
  \end{minipage}\hfill
  \begin{minipage}[t]{0.3\textwidth}
    \scriptsize
    \textbf{LeanHammer top-10:}
    \begin{enumerate}[leftmargin=1.4em, itemsep=0pt, nosep]
      \item \bpref{WeierstrassCurve.Affine.Point.map}
      \item \bpref{WeierstrassCurve.Affine.Point}
      \item \bpref{WeierstrassCurve.Affine.Point.map_injective}
      \item \bpref{WeierstrassCurve.Affine.Point.baseChange}
      \item \bpref{WeierstrassCurve.Affine.Point.instAddZeroClass}
      \item \bpref{WeierstrassCurve.Affine.Point.instZero}
      \item \bpref{WeierstrassCurve.Affine}
      \item \bpref{WeierstrassCurve.Affine.Point.instAddCommGroup}
      \item \bpref{Algebra}
      \item \bpref{WeierstrassCurve.toAffine}
    \end{enumerate}
    \smallskip
    \emph{Recall@32: \textbf{3/32} (0.09)}
  \end{minipage}\hfill
  \begin{minipage}[t]{0.3\textwidth}
    \scriptsize
    \textbf{ReProver top-10:}
    \begin{enumerate}[leftmargin=1.4em, itemsep=0pt, nosep]
      \item \bpref{ModuleCat.isFG}
      \item \bpref{ModuleCat.forget_obj}
      \item \bpref{LinearMap.id_fgModuleCat_comp}
      \item \bpref{FGModuleCat.hom_id}
      \item \bpref{CategoryTheory.Linear.fullSubcategory}
      \item \bpref{LinearMap.comp_id_fgModuleCat}
      \item \bpref{FGModuleCat.instFullModuleCatForget₂}
      \item \bpref{ModuleCat.hom_id}
      \item \bpref{ModuleCat.Algebra.instLinear}
      \item \bpref{FGModuleCat.carrier}
    \end{enumerate}
    \smallskip
    \emph{Recall@32: \textbf{0/32} (0.00)}
  \end{minipage}
  
  \medskip
  \textbf{Commentary.}
  \ourselector\ retrieves five gold premises in its top-10 
  (\bpref{Algebra.algebraMapSubmonoid}, 
  \bpref{Module.Basis.localizationLocalization}, 
  \bpref{Module.Basis}, 
  \bpref{Module.Basis.instFunLike}, 
  \bpref{Algebra.leftMulMatrix}) 
  and 11 of the 32 gold premises within its top-32. The remaining entries 
  cluster tightly in the \bpref{Algebra} and \bpref{Module.Basis} 
  namespaces that the query actually depends on. LeanHammer drifts to 
  \bpref{WeierstrassCurve.Affine.Point} (elliptic curve points), losing 
  the algebra and localization context entirely. ReProver drifts to 
  \bpref{ModuleCat} and \bpref{FGModuleCat} (the category-theoretic 
  recasting of module theory), retrieving zero gold premises in its 
  top-32. \ourselector's typed dependency edges resolve the precise 
  mathematical neighbourhood that this theorem consumes, while text-only 
  baselines latch onto unrelated parts of Mathlib that share only 
  superficial vocabulary.
  
  \end{tcolorbox}
  \caption{Case study from the \mlf-Random benchmark. \cmark\ marks 
  premises in the gold positive set. Premise names are Mathlib4 declaration 
  paths.}
  \label{fig:case-study-mathlib-random}
\end{figure*}

\FloatBarrier
\begin{figure*}[t]
  \centering
  \begin{tcolorbox}[
    colback=red!2!white, colframe=red!35, 
    coltitle=black,
    title={\textbf{Failure case study: \texttt{paul\_2\_mod\_5} 
           (\ourmethod-Collection)}},
    fontupper=\small
  ]
  
  \textbf{Query (Lean 4 statement):}
  \begin{lstlisting}[basicstyle=\ttfamily\footnotesize]
theorem paul_2_mod_5 (p : ℕ) (hp : p.Prime) (hp1 : p ≠ 5) :
    p % 5 ≠ 0 := by
  \end{lstlisting}
  
  \medskip
  \textbf{Gold positives (17, ranked by priority):}
  \bpref{Nat.instMulZeroClass}, \bpref{Nat.Prime}, \bpref{Nat.instOne}, 
  \bpref{Nat.instCharZero}, \bpref{Nat.Prime.eq_one_or_self_of_dvd}, 
  \bpref{Nat.instAddMonoidWithOne}, \bpref{AddGroup}, \bpref{neg_add_cancel}, 
  \bpref{add_neg_cancel}, \bpref{lt_trans}, \dots\ (17 total).
  
  \medskip
  
  \begin{minipage}[t]{0.3\textwidth}
    \scriptsize
    \textbf{\ourselector\ top-10:}
    \begin{enumerate}[leftmargin=1.4em, itemsep=0pt, nosep]
      \item \bpref{Rat.instNormedField}
      \item \bpref{Rat.nontrivial}
      \item \bpref{mul_ne_zero}
      \item \bpref{Rat.instCharZero}
      \item \bpref{Rat.instField}
      \item \bpref{Fact.elim}
      \item \bpref{NormMulClass.toNoZeroDivisors}
      \item \bpref{Rat.mul_den}
      \item \bpref{Real.instNontrivial}
      \item \bpref{Nat.Prime.ne_zero}
    \end{enumerate}
    \smallskip
    \emph{Recall@32: \textbf{0/17} (0.00)}
  \end{minipage}\hfill
  \begin{minipage}[t]{0.3\textwidth}
    \scriptsize
    \textbf{LeanHammer top-10:}
    \begin{enumerate}[leftmargin=1.4em, itemsep=0pt, nosep]
      \item \cmark\ \bpref{Nat.Prime}
      \item \bpref{Nat.decidablePrime}
      \item \bpref{Nat.Prime.ne_one}
      \item \bpref{Nat.Prime.two_le}
      \item \bpref{Nat.Prime.pos}
      \item \bpref{Nat.Prime.one_le}
      \item \bpref{Nat.Prime.one_lt}
      \item \bpref{Nat.Prime.ne_zero}
      \item \bpref{Nat.Prime.not_dvd_one}
      \item \bpref{Nat.Prime.prime}
    \end{enumerate}
    \smallskip
    \emph{Recall@32: \textbf{2/17} (0.12)}
  \end{minipage}\hfill
  \begin{minipage}[t]{0.3\textwidth}
    \scriptsize
    \textbf{ReProver top-10:}
    \begin{enumerate}[leftmargin=1.4em, itemsep=0pt, nosep]
      \item \bpref{Nat.Prime.ne_one}
      \item \cmark\ \bpref{Nat.Prime}
      \item \bpref{Nat.Prime.pos}
      \item \bpref{Nat.Prime.one_lt}
      \item \bpref{Nat.prime_def}
      \item \bpref{Nat.prime_def_lt}
      \item \bpref{Nat.Prime.ne_zero}
      \item \cmark\ \bpref{Nat.Prime.eq_one_or_self_of_dvd}
      \item \bpref{Nat.prime_def_lt'}
      \item \bpref{Nat.Prime.pred_pos}
    \end{enumerate}
    \smallskip
    \emph{Recall@32: \textbf{2/17} (0.12)}
  \end{minipage}
  
  \medskip
  \textbf{Commentary.}
  This is a failure mode for \ourselector. The query is a short fact about natural-number primality, but \ourselector's top-10 is dominated by typeclass instances on \bpref{Rat} and \bpref{Real} (\bpref{Rat.instNormedField}, \bpref{Rat.instField}, \bpref{Rat.instCharZero}, \bpref{Real.instNontrivial}) rather than the corresponding \bpref{Nat} instances that the proof actually rewrites against. \ourselector\ likely over-generalizes along the typeclass dimension. Several gold premises are typeclass instances that both \bpref{Nat} and \bpref{Rat} satisfy (\bpref{instCharZero}, \bpref{Field}, \bpref{IsStrictOrderedRing}), and \ourselector's structural features apparently collapse the two types into the same neighborhood, then pick the more densely-connected \bpref{Rat} variants. Text-only baselines avoid this confusion here because the query name \texttt{Nat.Prime} is lexically explicit. LeanHammer and ReProver both retrieve \bpref{Nat.Prime} and several \bpref{Nat.Prime.*} lemmas, and ReProver also surfaces the gold theorem \bpref{Nat.Prime.eq_one_or_self_of_dvd} at position 8. All three selectors still miss most of the typeclass infrastructure that the proof depends on, so absolute Recall@32 remains low for everyone. The case illustrates a real limitation of dependency-aware retrieval. When several types share the relevant typeclass structure, and the query carries a strong text signal toward only one of them, structural features alone can pick the wrong type.
  
  \end{tcolorbox}
  \caption{Failure case study from \ourmethod-Collection (a Lean Workbook 
  problem). \cmark\ marks premises in the gold positive set.
  \ourselector\ retrieves zero gold premises in its top-32 on this query, while LeanHammer and ReProver each retrieve two. Premise names are Mathlib4 declaration paths.
  }
  \label{fig:case-study-failure}
\end{figure*} 
\begin{figure*}[t]
  \centering
  \begin{tcolorbox}[
    colback=red!2!white,
    colframe=red!35,
    coltitle=black,
    title={\textbf{Failure case study: 
           \texttt{brownian-motion/lem:M\_uniform\_integrable} 
           (\ourmethod-Blueprint)}},
    fontupper=\small
  ]
  
  \textbf{Query (informal statement):}\\
  \textit{The sequence $(M^n_T)_{n \in \mathbb{N}}$ is uniformly integrable 
  (bounded in $L^1$ norm).}
  
  \medskip
  \textbf{Gold positives (2):}
  \bpref{brownian-motion/lem:A_uniform_integrable}, 
  \bpref{brownian-motion/def:classD}.
  
  \medskip
  
  \begin{minipage}[t]{0.3\textwidth}
    \scriptsize
    \textbf{\ourselector\ top-10:}
    \begin{enumerate}[leftmargin=1.4em, itemsep=0pt, nosep]
      \item \bpref{BonnAnalysis/lem:fourier-multiplication}
      \item \bpref{exchangeability/lem:covariance_structure}
      \item \bpref{brownian-motion/lem:uniformIntegrableIffNorm}
      \item \bpref{PrimeNumberTheoremAnd/RectangleIntegral_tendsTo_VerticalIntegral}
      \item \bpref{PrimeNumberTheoremAnd/ResidueTheoremOnRectangleWithSimplePole}
      \item \bpref{BonnAnalysis/thm:kernel-approximation}
      \item \bpref{PrimeNumberTheoremAnd/HolomorphicOn.vanishesOnRectangle}
      \item \bpref{brownian-motion/def:elemStochIntegralBilin}
      \item \bpref{exchangeability/lem:contractable_dist_eq}
      \item \bpref{brownian-motion/lem:uniformIntegrableDominatedSingleton}
    \end{enumerate}
    \smallskip
    \emph{Recall@32: \textbf{0/2} (0.00)}
  \end{minipage}\hfill
  \begin{minipage}[t]{0.3\textwidth}
    \scriptsize
    \textbf{LeanHammer top-10:}
    \begin{enumerate}[leftmargin=1.4em, itemsep=0pt, nosep]
      \item \bpref{brownian-motion/lem:uniformIntegrableNorm}
      \item \cmark\ \bpref{brownian-motion/lem:A_uniform_integrable}
      \item \bpref{brownian-motion/lem:Submartingale.uniformIntegrable_of_antitone_of_ge}
      \item \bpref{brownian-motion/lem:uniformIntegrableIffNorm}
      \item \bpref{brownian-motion/lem:uniformIntegrableDominatedSingleton}
      \item \bpref{brownian-motion/lem:uniformIntegrableAdd}
      \item \bpref{brownian-motion/def:HasIntegrableSup}
      \item \bpref{brownian-motion/lem:uniformIntegrableDominated}
      \item \bpref{radii-polynomial/thm:norm_mul_le}
      \item \bpref{brownian-motion/lem:covariation_eq_inner}
    \end{enumerate}
    \smallskip
    \emph{Recall@32: \textbf{1/2} (0.50)}
  \end{minipage}\hfill
  \begin{minipage}[t]{0.3\textwidth}
    \scriptsize
    \textbf{ReProver top-10:}
    \begin{enumerate}[leftmargin=1.4em, itemsep=0pt, nosep]
      \item \cmark\ \bpref{brownian-motion/lem:A_uniform_integrable}
      \item \bpref{brownian-motion/lem:uniformIntegrableNorm}
      \item \bpref{brownian-motion/lem:IsSquareIntegrable.sup_eLpNorm_eq_eLpNorm_limitProcess}
      \item \bpref{ClassFieldTheory/lem:norm_comm}
      \item \bpref{FLT_ch5/Hurwitz.norm}
      \item \bpref{brownian-motion/lem:Submartingale.uniformIntegrable_of_antitone_of_ge}
      \item \bpref{FLT_ch5/Hurwitz.norm_one}
      \item \bpref{FLT_ch5/Hurwitz.exists_near}
      \item \bpref{FLT_ch5/Hurwitz.norm_zero}
      \item \bpref{brownian-motion/cor:doob_lp_norm}
    \end{enumerate}
    \smallskip
    \emph{Recall@32: \textbf{1/2} (0.50)}
  \end{minipage}
  
  \medskip
  \textbf{Commentary.}
  This is a second failure mode for \ourselector, distinct from the failure in Appendix~\ref{fig:case-study-failure}. The query has only two gold premises. The first one, \bpref{A\_uniform\_integrable}, differs from the query name \bpref{M\_uniform\_integrable} by a single letter in the prefix, making it a strong lexical match that text-only baselines exploit directly. LeanHammer surfaces it at position 2 and ReProver at position 1. \ourselector\ instead drifts across the Blueprint corpus, retrieving thematically related premises about analysis and integrability from unrelated projects (\bpref{BonnAnalysis/lem:fourier-multiplication}, \bpref{PrimeNumberTheoremAnd/RectangleIntegral_tendsTo_VerticalIntegral}, \bpref{exchangeability/lem:covariance_structure}). \ourselector's structural features over-generalize across the global corpus when the query carries a strong lexical anchor that points to a tightly-named in-project sibling. Together with the type-level over-generalization illustrated in Appendix~\ref{fig:case-study-failure}, this case suggests that \ourselector\ would benefit from a stronger text-side anchoring signal to suppress structural neighbors that share thematic context but live in distant parts of the corpus.
  
  \end{tcolorbox}
  \caption{Failure case study from \ourmethod-Blueprint (the brownian-motion 
  project). \cmark\ marks premises in the gold positive set. 
  \ourselector\ retrieves zero gold premises in its top-32 here, while LeanHammer and ReProver each retrieve one (the lexically near-identical \texttt{A\_uniform\_integrable}). The case illustrates a second failure mode for \ourselector. When the gold premise shares strong lexical similarity with the query name, text-only baselines exploit the direct match, whereas \ourselector\ drifts across the global Blueprint corpus along thematic lines.}
  \label{fig:case-study-blueprint-failure}
\end{figure*} 
\begin{figure*}[t]
  \centering
  \begin{tcolorbox}[
    colback=red!2!white,
    colframe=red!35,
    coltitle=black,
    title={\textbf{Failure case study: \texttt{PrimeNumberTheoremAnd/ResidueOfTendsTo} 
           (\ourmethod-Blueprint)}},
    fontupper=\small
  ]
  
  \textbf{Query (informal statement):}\\
  \textit{If a function $f$ is holomorphic in a neighborhood of $p$ and 
  $\lim_{s \to p} (s - p)\,f(s) = A$, then 
  $f(s) = \dfrac{A}{s - p} + O(1)$ near $p$.}
  
  \medskip
  \textbf{Gold positives (1):}
  \bpref{PrimeNumberTheoremAnd/existsDifferentiableOn_of_bddAbove}.
  
  \medskip
  
  \begin{minipage}[t]{0.3\textwidth}
    \scriptsize
    \textbf{\ourselector\ top-10:}
    \begin{enumerate}[leftmargin=1.4em, itemsep=0pt, nosep]
      \item \bpref{PrimeNumberTheoremAnd/tendsto_rpow_atTop_nhds_zero_of_norm_gt_one}
      \item \bpref{PrimeNumberTheoremAnd/MellinConvolutionTransform}
      \item \bpref{radii-polynomial/prop:fixed_point_iff_zero}
      \item \bpref{PrimeNumberTheoremAnd/DeltaSpikeMass}
      \item \bpref{radii-polynomial/thm:neumann_series}
      \item \bpref{PrimeNumberTheoremAnd/MellinOfPsi}
      \item \bpref{radii-polynomial/prop:operator_norm_properties}
      \item \bpref{PrimeNumberTheoremAnd/MellinOfDeltaSpike}
      \item \bpref{radii-polynomial/lem:newton_fderiv}
      \item \bpref{carleson/nontangential-operator-boundary}
    \end{enumerate}
    \smallskip
    \emph{Recall@32: \textbf{0/1} (0.00)}
  \end{minipage}\hfill
  \begin{minipage}[t]{0.3\textwidth}
    \scriptsize
    \textbf{LeanHammer top-10:}
    \begin{enumerate}[leftmargin=1.4em, itemsep=0pt, nosep]
      \item \bpref{PrimeNumberTheoremAnd/isHolomorphicOn}
      \item \bpref{PrimeNumberTheoremAnd/HolomorphicOn_riemannZeta0}
      \item \bpref{PrimeNumberTheoremAnd/logDerivResidue}
      \item \bpref{PrimeNumberTheoremAnd/LogDerivZetaHolcSmallT}
      \item \bpref{PrimeNumberTheoremAnd/ResidueMult}
      \item \cmark\ \bpref{PrimeNumberTheoremAnd/existsDifferentiableOn_of_bddAbove}
      \item \bpref{PrimeNumberTheoremAnd/LogDerivZetaHolcLargeT}
      \item \bpref{PrimeNumberTheoremAnd/schwarz-id}
      \item \bpref{sphere-eversion/lem:param_trick}
      \item \bpref{PrimeNumberTheoremAnd/BddAbove_to_IsBigO}
    \end{enumerate}
    \smallskip
    \emph{Recall@32: \textbf{1/1} (1.00)}
  \end{minipage}\hfill
  \begin{minipage}[t]{0.3\textwidth}
    \scriptsize
    \textbf{ReProver top-10:}
    \begin{enumerate}[leftmargin=1.4em, itemsep=0pt, nosep]
      \item \bpref{FormalBook/ch26lemma_d}
      \item \bpref{M2/def:homotopy_diagram}
      \item \bpref{M2/def:homotopy_colimit}
      \item \bpref{FormalBook/ch41proof1}
      \item \bpref{FormalBook/ch41proof5}
      \item \bpref{FormalBook/ch41proof4}
      \item \bpref{FormalBook/ch41proof3}
      \item \bpref{FormalBook/ch41proof2}
      \item \bpref{Sphere-Packing-Lean/lemma:phi0-transform}
      \item \bpref{ABC-Exceptions/def:nu}
    \end{enumerate}
    \smallskip
    \emph{Recall@32: \textbf{0/1} (0.00)}
  \end{minipage}
  
  \medskip
  \textbf{Commentary.}
  This case illustrates a within-project topical drift for \ourselector. 
  The query is a residue-style statement about holomorphic functions, 
  and the unique gold premise is 
  \bpref{existsDifferentiableOn_of_bddAbove}, a technical lemma in 
  PNT's holomorphy/residue cluster. \ourselector\ stays inside the 
  \bpref{PrimeNumberTheoremAnd} and \bpref{radii-polynomial} projects 
  but drifts to a different topical subgraph, retrieving the Mellin-transform 
  and operator-norm lemmas (\bpref{MellinConvolutionTransform}, 
  \bpref{MellinOfPsi}, \bpref{MellinOfDeltaSpike}, 
  \bpref{neumann_series}, \bpref{operator_norm_properties}). LeanHammer's 
  text encoder identifies the correct PNT subcluster (\bpref{isHolomorphicOn}, 
  \bpref{HolomorphicOn_riemannZeta0}, \bpref{logDerivResidue}, 
  \bpref{ResidueMult}, \bpref{BddAbove_to_IsBigO}) and surfaces the gold 
  at position 6. ReProver scatters across unrelated projects 
  (\bpref{FormalBook}, \bpref{M2}, \bpref{Sphere-Packing-Lean}, 
  \bpref{ABC-Exceptions}) and also fails. The case shows that even 
  when \ourselector\ correctly localizes to the right project, its 
  structural features can pick the wrong topical subgraph within that 
  project when several thematically adjacent clusters compete.
  
  \end{tcolorbox}
  \caption{Failure case study from \ourmethod-Blueprint (the PrimeNumberTheoremAnd project). 
  \cmark\ marks premises in the gold positive set. 
  \ourselector\ retrieves zero gold premises in its top-32, while LeanHammer retrieves the single gold premise at position 6, and ReProver also fails. The case reveals a within-project topical drift for \ourselector. Structural features can resolve the right project but pick the wrong topical subgraph when several thematically adjacent clusters exist.}
  \label{fig:case-study-pnt-failure}
\end{figure*} 


\clearpage
\section*{NeurIPS Paper Checklist}

The checklist is designed to encourage best practices for responsible machine learning research, addressing issues of reproducibility, transparency, research ethics, and societal impact. Do not remove the checklist: {\bf The papers not including the checklist will be desk rejected.} The checklist should follow the references and follow the (optional) supplemental material.  The checklist does NOT count towards the page
limit. 

Please read the checklist guidelines carefully for information on how to answer these questions. For each question in the checklist:
\begin{itemize}
    \item You should answer \answerYes{}, \answerNo{}, or \answerNA{}.
    \item \answerNA{} means either that the question is Not Applicable for that particular paper or the relevant information is Not Available.
    \item Please provide a short (1--2 sentence) justification right after your answer (even for \answerNA). 
\end{itemize}

{\bf The checklist answers are an integral part of your paper submission.} They are visible to the reviewers, area chairs, senior area chairs, and ethics reviewers. You will also be asked to include it (after eventual revisions) with the final version of your paper, and its final version will be published with the paper.

The reviewers of your paper will be asked to use the checklist as one of the factors in their evaluation. While \answerYes{} is generally preferable to \answerNo{}, it is perfectly acceptable to answer \answerNo{} provided a proper justification is given (e.g., error bars are not reported because it would be too computationally expensive'' or ``we were unable to find the license for the dataset we used''). In general, answering \answerNo{} or \answerNA{} is not grounds for rejection. While the questions are phrased in a binary way, we acknowledge that the true answer is often more nuanced, so please just use your best judgment and write a justification to elaborate. All supporting evidence can appear either in the main paper or the supplemental material, provided in appendix. If you answer \answerYes{} to a question, in the justification please point to the section(s) where related material for the question can be found.

IMPORTANT, please:
\begin{itemize}
    \item {\bf Delete this instruction block, but keep the section heading ``NeurIPS Paper Checklist"},
    \item  {\bf Keep the checklist subsection headings, questions/answers and guidelines below.}
    \item {\bf Do not modify the questions and only use the provided macros for your answers}.
\end{itemize}


\begin{enumerate}

\item {\bf Claims}
    \item[] Question: Do the main claims made in the abstract and introduction accurately reflect the paper's contributions and scope?
    \item[] Answer: \answerYes{}
    \item[] Justification: The abstract and Section~\ref{sec:intro} state the data, method, benchmark, and downstream claims, and Sections~\ref{sec:dependency-resources}, \ref{sec:method}, and \ref{sec:experiments} report the corresponding evidence.

\item {\bf Limitations}
    \item[] Question: Does the paper discuss the limitations of the work performed by the authors?
    \item[] Answer: \answerYes{}
    \item[] Justification: Appendix~\ref{appx:limitations} the limitations. Appendix~\ref{appx:failure-cases} discusses three observed failure modes of \ourselector{} with detailed case studies.

\item {\bf Theory assumptions and proofs}
    \item[] Question: For each theoretical result, does the paper provide the full set of assumptions and a complete (and correct) proof?
    \item[] Answer: \answerNA{}
    \item[] Justification: The paper does not present formal theoretical theorems or proof claims. The mathematical content in Section~\ref{sec:method} defines the retrieval objective and training loss.

    \item {\bf Experimental result reproducibility}
    \item[] Question: Does the paper fully disclose all the information needed to reproduce the main experimental results of the paper to the extent that it affects the main claims and/or conclusions of the paper (regardless of whether the code and data are provided or not)?
    \item[] Answer: \answerYes{}
    \item[] Justification: Section~\ref{appx:reproducibility} summarizes reproducibility. Section~\ref{sec:data} introduces dataset construction. Section~\ref{sec:setup} and Appendix~\ref{appx:data-splits} describe the datasets, splits, baselines, and metrics. Appendix~\ref{appx:hyperparams} discloses training details and main hyperparameters.
    All dependency graphs, benchmarks, the extraction pipeline, and the model code are released at the URL in the abstract.

\item {\bf Open access to data and code}
    \item[] Question: Does the paper provide open access to the data and code, with sufficient instructions to faithfully reproduce the main experimental results, as described in supplemental material?
    \item[] Answer: \answerYes{}
    \item[] Justification: The data and code can be accessed via \url{https://daps-eth.github.io/}.

\item {\bf Experimental setting/details}
    \item[] Question: Does the paper specify all the training and test details (e.g., data splits, hyperparameters, how they were chosen, type of optimizer) necessary to understand the results?
    \item[] Answer: \answerYes{} 
    \item[] Justification: Training configuration and details are summarised in Table~\ref{tab:hyperparams}. Data splits and benchmark methodology are in Section~\ref{sec:data} and Appendix~\ref{appx:data-splits}.

\item {\bf Experiment statistical significance}
    \item[] Question: Does the paper report error bars suitably and correctly defined or other appropriate information about the statistical significance of the experiments?
    \item[] Answer: \answerNo{} 
    \item[] Justification: Main retrieval numbers in Table~\ref{tab:main-retrieval} are single-run point estimates due to compute budget. Per-bucket and per-category breakdowns in the analysis sections expose result stability through stratification rather than seeds. Variance across seeds remains a limitation.

\item {\bf Experiments compute resources}
    \item[] Question: For each experiment, does the paper provide sufficient information on the computer resources (type of compute workers, memory, time of execution) needed to reproduce the experiments?
    \item[] Answer: \answerYes{}
    \item[] Justification: Appendix~\ref{appx:compute} reports the hardware, device counts, runtime estimates, and aggregate compute for the training workloads used in the reported results.
    
\item {\bf Code of ethics}
    \item[] Question: Does the research conducted in the paper conform, in every respect, with the NeurIPS Code of Ethics \url{https://neurips.cc/public/EthicsGuidelines}?
    \item[] Answer: \answerYes{}
    \item[] Justification: The work uses public Lean libraries, public Blueprint projects, and public theorem proving benchmarks, and does not involve private personal data, human subjects, surveillance data, or safety-critical deployment.

\item {\bf Broader impacts}
    \item[] Question: Does the paper discuss both potential positive societal impacts and negative societal impacts of the work performed?
    \item[] Answer: \answerYes{}
    \item[] Justification: A brief broader impacts statement is included in Appendix~\ref{appx:broader-impacts}. The work bridges formal and informal mathematical reasoning by mining Lean Blueprints and improving LLM-based informal theorem proving, with no specific misuse concerns beyond general AI-for-mathematics considerations.
    
\item {\bf Safeguards}
    \item[] Question: Does the paper describe safeguards that have been put in place for responsible release of data or models that have a high risk for misuse (e.g., pre-trained language models, image generators, or scraped datasets)?
    \item[] Answer: \answerNA{}
    \item[] Justification: The paper releases theorem-proving graphs, benchmarks, and selector code rather than a high-risk generative model or scraped dataset with personal or sensitive content.

\item {\bf Licenses for existing assets}
    \item[] Question: Are the creators or original owners of assets (e.g., code, data, models), used in the paper, properly credited and are the license and terms of use explicitly mentioned and properly respected?
    \item[] Answer: \answerYes{} 
    \item[] Justification: Mathlib4 (Apache 2.0)~, the Lean Blueprint tool, and the source datasets for \ourmethod-Collection (\texttt{Goedel-LM/Lean-workbook-proofs}, \texttt{stoney0062/Leanabell-Prover-Traindata-SFT}, \texttt{AI-MO/NuminaMath-LEAN}) are cited with their respective licenses in Section~\ref{sec:data} and Appendix~\ref{appx:data-splits}.

\item {\bf New assets}
    \item[] Question: Are new assets introduced in the paper well documented and is the documentation provided alongside the assets?
    \item[] Answer: \answerYes{} 
    \item[] Justification: Four new evaluation benchmarks and the \ourselector{} model are introduced. Their construction, intended use, and license are documented in Section~\ref{sec:data}, Appendix~\ref{appx:data-splits}, and the README of the released repository (anonymous URL in abstract).

\item {\bf Crowdsourcing and research with human subjects}
    \item[] Question: For crowdsourcing experiments and research with human subjects, does the paper include the full text of instructions given to participants and screenshots, if applicable, as well as details about compensation (if any)? 
    \item[] Answer: \answerNA{}
    \item[] Justification: The paper does not involve crowdsourcing experiments or research with human subjects.

\item {\bf Institutional review board (IRB) approvals or equivalent for research with human subjects}
    \item[] Question: Does the paper describe potential risks incurred by study participants, whether such risks were disclosed to the subjects, and whether Institutional Review Board (IRB) approvals (or an equivalent approval/review based on the requirements of your country or institution) were obtained?
    \item[] Answer: \answerNA{}
    \item[] Justification: The paper does not involve crowdsourcing experiments or research with human subjects, so IRB approval is not applicable.

\item {\bf Declaration of LLM usage}
    \item[] Question: Does the paper describe the usage of LLMs if it is an important, original, or non-standard component of the core methods in this research? Note that if the LLM is used only for writing, editing, or formatting purposes and does \emph{not} impact the core methodology, scientific rigor, or originality of the research, declaration is not required.
    \item[] Answer: \answerYes{}
    \item[] Justification: Section~\ref{sec:informal} and Section~\ref{sec:downstream} describe the LLM use in the informal theorem proving experiment, including formalization with Gemini 3 Pro and proof generation by evaluated general-purpose LLMs.

\end{enumerate}

\end{document}